\documentclass[11pt,draftclsnofoot,journal,onecolumn]{IEEEtran}
\usepackage{graphicx}
\usepackage{epstopdf}
\usepackage[latin1]{inputenc}
\usepackage{amsmath,amssymb,amscd,latexsym,dsfont}
\usepackage[english]{babel}
\usepackage{tabularx,cite}
\usepackage{graphicx}
\usepackage{algpseudocode}
\usepackage{algorithm}
\usepackage{amsmath,bm}
\usepackage{algorithmicx}
\usepackage{float}

\usepackage{multicol}
\usepackage{graphicx}
\usepackage{caption}
\usepackage{psfrag}
\usepackage{epstopdf}
\usepackage{comment,psfrag,subfigure,enumerate}
\usepackage{color}
\usepackage{amsthm}
\usepackage{bigints}
\usepackage{lipsum}
\usepackage{relsize}
\usepackage{graphicx}
\usepackage{epstopdf}
\usepackage{calligra}
\theoremstyle{definition}
\usepackage{mathtools}
\newtheorem{theorem}{Theorem}

\newtheorem{coro}{Corollary}

\ifCLASSINFOpdf
\else
\fi
\begin{document}
%
\title{On Wireless Energy and Information Transfer in Relay Networks}
\author{\IEEEauthorblockN{Mahdi Haghifam, Behrooz Makki, Masoumeh Nasiri-Kenari and Tommy Svensson}\\
\thanks{Mahdi Haghifam and Masoumeh Nasiri-Kenari are with Sharif University of Technology, Tehran, Iran, Email: haghifam$\_$mahdi@ee.sharif.edu, mnasiri@sharif.edu. Behrooz Makki and Tommy Svenesson are with Chalmers University of Technology, Gothenburg, Sweden, Email: \{behrooz.makki, tommy.svensson\}@chalmers.se. }
\thanks{This work partially presented at the IEEE PIMRC 2016.}
}

%
\maketitle
\vspace{-16mm}
\begin{abstract}
This paper investigates the outage probability and the throughput of relay networks with wireless information and energy transfer where the relays harvest energy from the transmitted radio-frequency signal of the source. Considering different power consumption models, we derive the outage probability for both adaptive and non-adaptive power allocations at the relay. With a total energy consumption constraint at the source, we provide closed-form expressions for the optimal time sharing and power allocation between the source energy and information transfer signals as well as the optimal relay positioning such that the outage probability is minimized. Finally, we extend our analysis to multi-relay networks. We show that with perfect channel state information (CSI) available at the relays and $N$ relays the opportunistic relaying scheme achieves diversity order of $\frac{N+1}{2}$. Also, we analyze the opportunistic relaying with partial CSI where either the source-relay or the relay-destination CSI is provided at its corresponding transmit terminal, and prove that the relay selection based on the source-relay CSI outperforms the relay selection based on the relay-destination CSI, in terms of outage probability. The analytical and simulation results demonstrate the efficiency of wireless energy and information transfer systems in different conditions.
\end{abstract}
%
\IEEEpeerreviewmaketitle
\vspace{-10mm}
\section{Introduction}
One of the promising methods to ensure high quality of service in wireless networks is to use many small cheap nodes that support information transfer between the terminals. These devices are usually powered by fixed but limited batteries. Thus, wireless networks may suffer from short lifetime and, to prolong their lifetime, they require periodic battery replacement/recharging. However, the battery replacement may be infeasible in, e.g., biological or chemical environments. Energy harvesting is a promising solution for such problems where the nodes harvest energy from, e.g., wind, solar, kinetic sources\cite{tutoreh}. The main drawback of such solution is that their performance depend much on the weather condition. For this reason, it has been recently proposed to use radio-frequency (RF) signals as a means of wireless energy transfer. Significant advances in the circuit design for RF energy transfer make the usage of energy transfer as a viable and practical solution for future wireless networks, e.g., \cite{tutnew1,tutorwet,tutnew2}.\par
From another prospective, relay-assisted communication is one of the promising techniques
that have been proposed for wireless networks.
The main idea of a relay network is to improve data
transmission efficiency by implementation of intermediate
relay nodes which support data transmission from a source
to a destination. The relay networks have been adopted in the
3GPP long-term evolution advanced (LTE-A) standardization
\cite{relay_beh} and are expected to be one of the core technologies for
the next generation cellular systems. These are the motivations for this paper analyzing the performance of the relay networks with wireless energy transfer.
\par
The concept of using a RF signal for simultaneously wireless information and energy transfer is introduced in \cite{alaki1,alaki2} for flat and frequency selective fading channels, respectively. Practical receiver design and modulation techniques for wireless information and energy transfer are studied in \cite{cirs,cirm} for single-input-single-output (SISO) and multiple-input-multiple-output (MIMO) channels, respectively. In \cite{13,10}, the outage-optimized power control policies for the energy harvesting transmitters are investigated for finite and infinite time horizon, respectively. Also, \cite{encoop} studies the single link network with cooperative energy harvesting transmitter and receiver. Some recent studies \cite{10,savetrans,encoop} deal with energy consumption model for the single link energy harvesting networks. 
\par
Relay networks with simultaneous wireless energy and information transfer are further studied in \cite{rele1,rele2,rele3}. Different harvest-then-transmit and harvest-then-cooperate protocols for cooperative communication in networks with energy transfer are introduced in \cite{thrmax} and \cite{harvtran}, respectively. Then, \cite{naz} analyzes the ergodic and outage capacities of the decode-and-forward relays with co-channel interference. In \cite{nasir13,nasir15}, relaying protocols with simultaneous wireless information and energy transfer are proposed. Moreover,\cite{poor} derives different power allocation strategies for energy harvesting relay networks with multiple source-destination pairs and a single energy harvesting relay. The multi-relay network with information and energy transfer are also studied in \cite{100,poor_rand,maxmin_alaki,kiri_finite}. Recently, \cite{tars} studies the Max-Min scheduling in multi-user cooperative networks with a single relay that utilizes the power splitting method.\par
In this paper, we study the performance of relay networks with wireless information and energy transfer and non-ideal (realistic)
assumption on the power consumption of the relay. The design problem is cast in the form of minimizing the outage
probability subject to a total energy consumption constraint at the source. We use the a switching protocol for the relay-based data/energy transfer. Our contributions are as follows:
\begin{enumerate}
\item We derive closed-form expressions for the optimal time sharing between the energy and information signals such that the  energy-limited outage probability is minimized (Theorem \ref{thr:tet}). Also, we obtain closed-form expressions for the optimal, in terms of energy-constrained outage probability, power allocation of the source (Theorem \ref{thr:pow}). Moreover, we study the optimal position of the relay that minimizes the outage probability. Finally, we investigate the system performance in the cases with perfect CSI at the source as a benchmark for other schemes based on the channel distribution information (CDI).
\item For the multi-relay network with $N$ relays, we consider the opportunistic relaying with perfect CSI and derive the outage probability and high signal-to-noise-ratios (SNR) performance of the system. Subsequently, it is shown that the diversity order of the scheme is $\frac{N+1}{2}$ (Theorem \ref{thr:opor_full} and Corollary \ref{corr:opor_full}).
\item  We also evaluate the outage performance of the opportunistic relaying in the cases where either the source-relay or the relay-destination CSI is available, and show that in this case the diversity order is equal to 1. We prove that the relay selection criteria based on the source-relay CSI has better,  in terms of outage probability, performance than the one based on the relay-destination CSI  (Theorem \ref{thr:opor_partial} and Corollary \ref{corr:opor_partial}).   
\end{enumerate}
 \par
Compared to the literature, e.g., \cite{10,maxmin_alaki,100,kiri_finite,harvtran,rele1,rele2,rele3,thrmax,tars,savetrans,encoop,poor,nasir13,nasir15,naz,poor_rand}, we consider different power amplifier (PA) model, optimization criteria/metrics, and problem formulation, which lead to completely different analysis/conclusions. Moreover, our discussions on the optimal time sharing between energy and information signals, the optimal relay position, the optimal power allocation at the source, diversity order of opportunistic relaying in energy harvesting networks, 
and comparison of different relay selections criteria based on the available CSI have not been presented before.\par
Our analytical results which have been confirmed by simulations indicate that the optimal time sharing, with respect to the source transmission power, has two regions in which the optimal time sharing is independent and increasing functions of the source transmit power (Theorem \ref{thr:tet} and Fig. \ref{fig:optimal_theta}). Furthermore, at high SNRs, the optimal power for the energy transfer signal, in terms of outage probability, increases linearly with the total energy constraint of the source and decreases exponentially with the codeword rate (Theorem \ref{thr:pow}, Eq. (\ref{eq:opt_p_hsnr}), and Fig. \ref{fig:opt_pow_1}). Finally, considering  $N$ relays in a multi-relay networks with perfect and partial CSI, the outage probability of the opportunistic relaying increases with the inefficiency of the PA in power
of $\frac{N+1}{2}$ and $1$, respectively (Corollaries \ref{corr:opor_full} and \ref{corr:opor_partial}).\par 
The outline of this paper is as follows. In Section \ref{sec:sysmod}, the system model is described. In Sections \ref{sec:non} and \ref{sec:adp}, we analyze the system performance for the single relay networks with the non-adaptive and adaptive power allocation at the relay, respectively. Then, in Section \ref{sec:multi}, the multi-relay network is analyzed. The simulation results are given in Section \ref{nemodara}, where we verify the analytic results. Finally, Section \ref{sec:conc} concludes the paper.
\section{SYSTEM MODEL} \label{sec:sysmod}
We consider a relay-assisted cooperative communication setup consisting of a source, a relay, and a destination. In harmony with, e.g., \cite{paperb1,paperb2,paperb3,nasir13,nasir15,tars}, we ignore the direct link between the source and the destination. The source and the destination nodes have constant, e.g., wired, power supply. On the other hand, the relay has no fixed power supply and receives its required energy from the source wirelessly. 
The channel coefficients in the source-relay and the relay-destination links are denoted by $h_{\text{sr}}$ and $h_{\text{rd}}$, respectively. The channel coefficients remain constant during the channel
coherence time and then change according to their probability distribution functions (PDFs). The source-relay and the relay-destination distances are denoted by $d_{\text{sr}}$ and $d_{\text{rd}}$, respectively. The expected channel gain is modeled as $\lambda_{\vartheta}=\alpha_{\vartheta}d_{\vartheta}^{-\beta_{\vartheta}}$ for $\vartheta=\{\text{sr},\text{rd}\}$, where $\beta_{\vartheta}$ is the path loss exponent of the corresponding link and $\alpha_{\vartheta}$ is a parameter independent of $d_{\vartheta}$ and determined according to, e.g., the transmitter and the receiver antenna gains and shadowing. Also, we define the channel gains as $g_{\text{sr}}=|h_{\text{sr}}|^2$ and $g_{\text{rd}}=|h_{\text{rd}}|^2$. The results are obtained for Rayleigh fading channels where the channel gains PDF are given by $f_{\vartheta}\left(x\right)=\frac{1}{\lambda_{\vartheta}}\exp\left(-\frac{x}{\lambda_{\vartheta}}\right)$ for $\vartheta=\{\text{sr},\text{rd}\}$.\par
Let us denote the total packet transmission length from the source to the destination by $T$. The energy transfer and data communication protocol is in three phases as follows.
In the first phase, of length $(1-\theta)T, \theta\in[0,1],$ the relay harvests energy from the source transmitted energy signal. Let $P_\text{s}^{\text{e}}$ be the transmission power of the source during the energy transmission phase. Then, the baseband signal model in this period is given by
\begin{equation} \label{eq:bas_chan_en}
y^{\text{e}}_{\text{r}}=\sqrt{P_{\text{s}}^{\text{e}}}h_{\text{sr}} + z^{\text{e}}_{\text{r}},
\end{equation}
where $z^{\text{e}}_{\text{r}}$ is the additive white Gaussian noise (AWGN) of the energy receiver of the relay with variance $\left(\sigma^{\text{e}}_\text{r}\right)^2$. For simplicity, we set $\sigma^{\text{e}}_\text{r}=0$. This is motivated by the fact that in many practical energy harvesting systems the harvested energy due to the noise (including both the antenna noise and the rectifier noise) is negligible, e.g., \cite{cirs}. Also, it is straightforward to extend the results to the cases with different noise variances. In this way, the energy harvested by the relay at the end of the energy transfer phase is given by $E_\text{stored}=(1-\theta)T\eta g_\text{s,r}P_\text{s}^\text{e}$, with $\eta$ representing the efficiency factor of the energy harvesting circuit. With no loss of generality, we set $\eta=1$. Also, we assume an ideal battery for the relay, such that no overflow occurs. Also, it is straightforward to extend the results to the cases with a peak energy storage for the battery. Let us define $E_\text{proc}$ as the minimum energy required by the relay to process the source signal and send feedbacks. Once the relay's required energy, $E_\text{proc}$, is supplied in Phase 1, it sends one bit acknowledgement to the source, and the information transfer starts where the information is forwarded to the destination through the relay (the consumed energy for sending one bit feedback is included in the relay's minimum required energy).\\
Receiving the acknowledgement from the relay, the second phase of length $\frac{\theta}{2}T$ starts where the source sends information to the relay. Let $x^{\text{i}}_{\text{s}} \in \mathcal{CN}(0,1)$ and $y_\text{r}^\text{i}$ be the source information signal and its corresponding received signal by the relay, respectively. Hence the channel is modeled as
\begin{equation}\label{eq:bas_chan_it1}
y^{\text{i}}_{\text{r}}=\sqrt{P_{\text{s}}^{\text{i}}} h_{\text{sr}}x_{\text{s}}^{\text{i}} + z_{\text{r}}^{\text{i}},
\end{equation}
with $P_\text{s}^\text{i}$ denoting the source power for the information signal, and $z_{\text{r}}^{\text{i}}$ is the AWGN of the information receiver of the relay with variance $\left(\sigma_{\text{r}}^{\text{e}}\right)^2$.\\
Finally, if the relay correctly decodes the source message, it uses the last time slot of length $\frac{\theta}{2}T$ to forward the codeword to the destination.
Thus, the destination's received signal is given by
\begin{equation}\label{eq:bas_chan_it2}
y_{\text{d}}=\sqrt{P_{\text{r}}^{\text{i}}}h_{\text{rd}}x_{\text{r}} + z_{\text{d}},
\end{equation}
where the relay signal $x_{\text{r}}$ follows $\mathcal{CN}(0,1)$,  $z_{\text{d}}$ is the additive white Gaussian noise with with variance $\sigma_{\text{d}}^2$ and $P_{\text{r}}^{\text{i}}$ is the relay's information transmission power. Finally, with no loss of generality, we set $\sigma_{\text{r}}^{\text{e}}=\sigma_{\text{d}}=1$ and assume the packets to be sufficiently long such that the results are independent of $T$, so it can be removed from the analysis.\par
We assume an ideal energy consumption model for the source, motivated by the fact that the base stations are commonly equipped with considerably stronger PAs than the relays. In the meantime, it is straightforward to extend the results to the cases with non-ideal PAs at the source. On the other hand, we adopt an affine model for the power consumption of the relay, where the relay's power consumption is modeled by
\begin{equation}\label{eq:power_cons_model}
P_{\text{cons}}=P_{\text{active}} + \nu P_{\text{r}}^{\text{i}}.
\end{equation}
Here, $\nu \geq 1$ represents the inefficiency of the PA and $P_\text{active}$ represents the relay's bias power during data transmission. This is a well-established model for the PAs \cite{paperjin1,paper_AB} and has been considered for different applications \cite{10,paperjin2}.\par
We analyze the system performance for two power allocation models at the relay:
\begin{enumerate}
\item \textbf{Non-adaptive transmission power}. Here, the relay has a predefined (peak) transmission power. This is an appropriate assumption in the cases with simple relays that can not adopt the transmission power.
\item \textbf{Adaptive transmission power}. Here, the relay adaptively updates its transmission power to forward the data to the destination with the maximum possible power.
\end{enumerate}
In Sections \ref{sec:non} and \ref{sec:adp}, we analyze the system outage probability for the cases with non-adaptive and adaptive power allocation models at the relay, respectively.
\vspace{-3mm}
\section{Performance Analysis For Non-adaptive Relays} \label{sec:non}
Let $E_{\text{proc}}=\frac{\theta}{2}P_{\text{proc}}$ where $P_{\text{proc}}$ is an auxiliary variable to simplify the analytical expressions. With a non-adaptive relay, a constant power $P_\text{r}^{\text{i}}$ is used by the relay to forward the information to the destination. Thus, from (\ref{eq:power_cons_model}), the total energy consumed by the relay during the second and the third phases is
\begin{equation}
E_{\text{r}}=\frac{\theta}{2}(P_{\text{proc}}+P_{\text{cons}})=\frac{\theta}{2}\left(P_{\text{proc}} + P_{\text{active}}+ \nu P_\text{r}^{\text{i}}\right).
\end{equation}  
In this way, if the the relay's minimum energy is not supplied in Phase 1, the relay does not become active, and a circuit outage event occurs. Thus, the probability of circuit outage for non-adaptive relay is
\begin{equation} \label{eq:cc_oo_ff}
\begin{aligned}
\mathrm{Pr}(\text{Circuit Outage})&= \mathrm{Pr}\left( g_{\text{sr}}P_{\text{s}}^{\text{e}}(1-\theta) < \frac{\theta}{2}(P_{\text{cons}}+P_{\text{proc}})\right)
&=1-\exp\left(-\frac{(P_{\text{cons}}+P_{\text{proc}}) \frac{\theta}{2}}{\lambda_{\text{sr}} P_{\text{s}}^{\text{e}}(1-\theta)}\right).
\end{aligned}
\end{equation}
Here, the last equality holds for Rayleigh fading conditions where the source-relay channel gain $g_{\text{sr}}$ follows an exponential distribution with mean $\lambda_{\text{sr}}$.
Define the effective SNR as
\begin{equation} \label{eq:DF_SNR}
\gamma_{\text{eff}}=\min\{\gamma_{\text{sr}},\gamma_{\text{rd}}\},
\end{equation}
where $\gamma_{\text{sr}}=g_{\text{sr}}P^{\text{i}}_{\text{s}}$ and $\gamma_{\text{rd}}=g_{\text{rd}}P_{\text{r}}^{\text{i}}$ are the relay's and the destination's SNRs, respectively.
In this way, representing the code rate by $R$, the rate outage probability, i.e., the probability that the data is not correctly decoded by the relay or the destination given that the relay required energy is supplied, is found as
\begin{equation} \label{eq:prob_ro_no}
\begin{aligned}
&\mathrm{Pr}(\text{Rate Outage}\big|\text{No Circuit Outage})
=1 - \mathrm{Pr}\left(\frac{\theta}{2}\log\left(1+\gamma_{\text{eff}}\right) \geq R \big|\text{No Circuit Outage}\right)\\&= 1 - \mathrm{Pr}\left( g_{\text{sr}} P_\text{s}^\text{i} \geq \gamma ,g_{\text{rd}} P_{\text{r}}^{\text{i}} \geq \gamma \big| g_{\text{sr}} \geq \frac{\frac{\theta}{2}(P_{\text{cons}}+P_{\text{proc}})}{P_{\text{s}}^{\text{e}}(1-\theta)}\right)\\
&= 1 - \mathrm{Pr}\left( g_{\text{sr}} P_{\text{s}}^{\text{i}} \geq \gamma \big| g_{\text{sr}} \geq \frac{\frac{\theta}{2}(P_{\text{cons}}+P_{\text{proc}})}{P_{\text{s}}^{\text{e}}(1-\theta)}\right)\mathrm{Pr}\left(g_{\text{rd}} P_{\text{r}}^{\text{i}}\geq \gamma \right)\\
&=\begin{cases}\vspace{0.3cm}
1-\exp\left(-\frac{\gamma}{\lambda_{\text{rd}} P_{\text{r}}^{\text{i}}}\right) &\text{$ \frac{P_{\text{proc}}+P_{\text{cons}}}{P_{\text{s}}^{\text{e}}(1-\theta)} \geq \frac{\gamma}{P_{\text{s}}^{\text{i}}\frac{\theta}{2}} $}\\ 
1-\frac{\exp\left(-\frac{\gamma}{\lambda_{\text{sr}} P_{\text{s}}^{\text{i}}}\right)\exp\left(-\frac{\gamma}{\lambda_{\text{rd}} P_{\text{r}}^{\text{i}}}\right)}{\exp\left(-\frac{\frac{\theta}{2} (P_{\text{cons}}+P_{\text{proc}})}{\lambda_{\text{sr}} P_{\text{s}}^{\text{e}}(1-\theta)}\right)} &\text{ $ \frac{P_{\text{proc}}+P_{\text{cons}}}{P_{\text{s}}^{\text{e}}(1-\theta)} \leq \frac{\gamma}{P_{\text{s}}^{\text{i}}\frac{\theta}{2}}  $}
\end{cases},
\end{aligned}
\end{equation}
where, in the last equality, $\gamma=\exp\left(\frac{2R}{\theta}\right)-1$ is an auxiliary variable to simplify the expressions. \\
 Using (\ref{eq:cc_oo_ff}), (\ref{eq:prob_ro_no}) and some manipulations, the outage probability for the non-adaptive power allocation is found as
\begin{equation} \label{eq:ov_out_f}
\begin{aligned}
\mathrm{Pr}(\text{outage}) &= \mathrm{Pr}(\text{Circuit Outage}) +  \mathrm{Pr}(\text{Rate Outage} | \text{No Circuit Outage}) \mathrm{Pr}(\text{No Circuit Outage})\\  
&=\begin{cases}\vspace{0.3cm}
1- \exp\left(-\frac{(P_{\text{cons}}+P_{\text{proc}}) \frac{\theta}{2} }{\lambda_{\text{sr}} P_{\text{s}}^{\text{e}}(1-\theta)}\right)\exp\left(-\frac{\gamma}{\lambda_{\text{rd}} P_{\text{r}}^{\text{i}}}\right) &\text{ $ \frac{P_{\text{proc}}+P_{\text{cons}}}{P_{\text{s}}^{\text{e}}(1-\theta)} \geq \frac{\gamma}{P_{\text{s}}^{\text{i}}\frac{\theta}{2}}$}\\ 
1- \exp\left(-\frac{\gamma}{\lambda_{\text{sr}} P_{\text{s}}^{\text{i}}}\right)\exp\left(-\frac{\gamma}{\lambda_{\text{rd}} P_{\text{r}}^{\text{i}}}\right) &\text{ $\frac{P_{\text{proc}}+P_{\text{cons}}}{P_{\text{s}}^{\text{e}}(1-\theta)} \leq \frac{\gamma}{P_{\text{s}}^{\text{i}}\frac{\theta}{2}}$}
\end{cases}.
\end{aligned}
\end{equation}
It is worth noting that in the high-SNR regime, the outage in (\ref{eq:ov_out_f}) is given by
\begin{equation}\label{eq:shirr}
\mathrm{Pr}\left(\text{outage}\right) \simeq 1 - \exp\left(-\frac{\exp\left(\frac{2R}{\theta}\right)-1}{\lambda_{\text{rd}} P_{\text{r}}^{\text{i}}}\right).
\end{equation}
Finally, the source expected consumed energy is obtained by
\begin{equation}\label{eq:const_power_f}
\begin{aligned}
\bar E_\text{s}&= P_{\text{s}}^{\text{e}} (1-\theta)+ P_{\text{s}}^{\text{i}}\frac{\theta}{2} \mathrm{Pr} (\text{source becomes active in Phase 2})\\
&=P_{\text{s}}^{\text{e}}(1-\theta)+ P_{\text{s}}^{\text{i}} \frac{\theta}{2} \exp\left(-\frac{(P_{\text{cons}}+P_{\text{proc}}) \frac{\theta}{2} }{\lambda_{\text{sr}} P_{\text{s}}^{\text{e}}(1-\theta)}\right).
\end{aligned}
\end{equation}
 Thus, considering a total energy constraint $\bar E_\text{s}\le E_{\text{max}}$, the energy-constrained outage minimization problem can be written as 
\begin{equation} \label{opt_p:tet}
\begin{aligned}
& \underset{ \theta,P_{\text{s}}^{\text{i}},P_{\text{s}}^{\text{e}}, d_{\text{sr}}, d_{\text{rd}}}{\text{minimize}}
& & \mathrm{Pr}(\text{outage}) 
& \text{subject to}
& & \bar E_{\text{s}}\leq E_{\text{max}}=P_{\text{max}}T,
& & d_{\text{sr}}+d_{\text{rd}}=d,\;
\end{aligned}
\end{equation}
with $P_\text{s}^\text{e}, P_\text{s}^\text{i}, \theta, d_{\text{sr}},$ and $d_{\text{rd}}$ being the optimization parameters and $d$ denotes the source-destination distance.  Since, we normalize $T=1$, the parameters $E_{\text{max}}$ and $P_{\text{max}}$ can be used interchangeably. In the following, we optimize the time sharing between the information and energy signals, the power allocations at the source as well as the relay's position such that the energy-constrained outage probability is minimized. 
\subsection{Optimal Time Sharing}
In this subsection, we minimize the outage probability in (\ref{opt_p:tet}) with respect to $\theta$ in the cases with a peak power constraint at the source, i.e., $P_\text{s}^\text{e}=P_\text{s}^\text{i}\leq P_0$, which is solved in Theorem 1.
\begin{theorem}\label{thr:tet} Optimal time sharing between the power and information signals is given by
\end{theorem}
\vspace{-25pt}
\begin{align} \label{eq:opt_thettttt}
\theta^{*}=
\begin{cases}\vspace{0.1cm}
\theta^{*}_1& \frac{2(1-\theta^{*}_1)\left(\exp\left(\frac{2R}{\theta^{*}_1}\right)-1\right)}{\theta^{*}_1} \leq P_{\text{cons}}+P_{\text{proc}}\\ 
\theta^{\star}_2&\text{O.W}
\end{cases},
\end{align}
where $\theta^{*}_1=\frac{R}{R + \mathcal{W}\left(\frac{\sqrt{\lambda_{\text{rd}}P_{\text{r}}^{\text{i}}(P_{\text{cons}}+P_{\text{proc}})R}}{2\sqrt{\lambda_{\text{sr}}P_0}\exp(R)}\right)}$, $\theta^{\star}_2\simeq\frac{2R}{2R+\mathcal{W} \left( R(P_{\text{cons}}+P_{\text{proc}})\exp(-2R)\right)}$, and $\mathcal{W}(x)$ denotes the Lambert $\mathcal{W}$ function \cite{onlambert}.
\begin{proof}
See Appendix A.
\end{proof}
Although the optimal point in the lower branch of (\ref{eq:opt_thettttt}) is derived for moderate/large codeword rates, Fig. \ref{fig:optimal_theta} indicates that the approximation is very tight for broad ranges
of codeword rates, because in (\ref{eq:fixed_pp}), we have $\exp\left(\frac{2R}{\theta}\right)\gg 1$ even for small $R$'s. Moreover, the optimal time sharing is independent of $P_{0}$ in the lower branch of (\ref{eq:opt_thettttt}). Finally, at high SNRs, since $\lim_{x \to 0} \mathcal{W}\left(x\right)=0$, the optimal time sharing in upper branch of (\ref{eq:opt_thettttt}) converges into 1, as expected. 
\subsection{Optimal Power Allocation}
Considering given signals lengths and the energy-limited outage minimization problem, i.e., $\theta$, the optimal power allocation at the source between energy and information signals is provided in Theorem \ref{thr:pow}. 
\begin{theorem}\label{thr:pow} Optimal, in terms of outage probability in (\ref{eq:ov_out_f}), source power allocation for the energy and information transfer signals is given by
\end{theorem}
\vspace{-17pt}
\begin{equation}\label{opt_point_22}
\begin{aligned}
&(P_{\text{s}}^{\text{e}})^{\star}=\frac{P_{\text{max}}\frac{\theta}{2}\left(P_{\text{proc}}+P_{\text{cons}}\right)}{\lambda_{\text{sr}}P_{\text{max}}\left(1-\theta\right)\mathcal{W}\left(\frac{\gamma\theta}{\lambda_{\text{sr}}P_{\text{max}}}\exp\left(-\frac{\left(P_{\text{proc}}+P_{\text{cons}}\right)\frac{\theta}{2}}{\lambda_{\text{sr}}P_{\text{max}}}\right)\right)+\left(P_{\text{proc}}+P_{\text{cons}}\right)\left(1-\theta\right)\theta},\\
&(P_{\text{s}}^{\text{i}})^{\star} = \frac{P_{\text{max}}-(P_{\text{s}}^{\text{e}})^{\star}(1-\theta)}{\frac{\theta}{2} \exp\left(-\frac{1}{\lambda_{\text{sr}}}\frac{P_{\text{proc}}+P_{\text{cons}}}{(P_{\text{s}}^{\text{e}})^{\star}}\frac{\frac{\theta}{2}}{1-\theta}\right)},
\end{aligned}
\end{equation}
where $\mathcal{W}\left(x\right)$ is the Lambert W function \cite{onlambert}. \begin{proof}
See Appendix B.
\end{proof}
Using (\ref{opt_point_22}) and $\lim_{x \to 0^{+}}\mathcal{W}'\left(x\right)=1$, for $P_{\text{max}}\rightarrow \infty$ \cite{onlambert}, we have
\begin{equation}\label{eq:opt_p_hsnr}
(P_{\text{s}}^{\text{e}})^{\star}\simeq \left(\frac{\left(P_{\text{proc}}+P_{\text{cons}}\right)}{2\left(1-\theta\right)\left(\exp\left(\frac{2R}{\theta}\right)+P_{\text{proc}}+P_{\text{cons}}\right)}\right)P_{\text{max}},
\end{equation}
which implies that, the optimal power for energy transfer increases linearly with $P_{\text{max}}$ and also decreases exponentially with the codeword rate.
In Fig. \ref{fig:opt_pow_1}, we verify the analytical results of Theorem \ref{thr:pow} by comparing them with the simulation results.
\subsection{Optimal Relay Position}
In this section, we study the optimal position of the relay such that the energy-constrained outage probability is minimized. We consider a setup where the relay's position can be changed on the line between the source and the destination. Also, we assume $\alpha_{\text{sr}}=\alpha_{\text{rd}}=\alpha$ and $\beta_{\text{sr}}=\beta_{\text{rd}}=\beta \geq 1$ which is a reasonable assumption in the relay-assisted communication networks. For simplicity, we assume that if the relay is placed in the midpoint of the source-destination, the expected channel gains of the source-relay and the relay-destination links are equal to one. Thus, we have $\alpha d^{-\beta}=2^{-\beta}$. 
For mathematical convenience, we define $\delta=\frac{d_{\text{sr}}}{d_{\text{rd}}}$. Thus, $d_{\text{sr}}=\frac{d\delta}{1+\delta}$ and $d_{\text{rd}}=\frac{d}{1+\delta}$. From (\ref{eq:const_power_f}), we have the following constraint on $\delta$
\begin{equation} \label{eq:const_delta}
\begin{aligned}
&P_{\text{s}}^{\text{e}}(1-\theta)+ P_{\text{s}}^{\text{i}} \frac{\theta}{2} \exp\left(-\frac{(P_{\text{cons}}+P_{\text{proc}}) \frac{\theta}{2} }{\lambda_{\text{sr}} P_{\text{s}}^{\text{e}}(1-\theta)}\right) \leq P_{\text{max}}\\
&\Rightarrow  \delta \geq \frac{\sqrt[\beta]{A}}{1-\sqrt[\beta]{A}}, 
 \ \ \ A\triangleq 2^{-\beta}\frac{P_{\text{s}}^{\text{e}}(1-\theta)}{(P_{\text{cons}}+P_{\text{proc}})\frac{\theta}{2}}\log\left(\frac{P_{\text{s}}^{\text{i}}\frac{\theta}{2}}{P_{\text{max}}-P_{\text{s}}^{\text{e}}(1-\theta)}\right) ,
\end{aligned}
\end{equation}
if $A\geq 0$, otherwise, every $\delta\ge0$ is the feasible set. Also, it is straightforward to prove that $A<1$. Therefore, for every given powers $P_\text{proc}, P_\text{cons},P_{\text{s}}^{\text{i}},P_{\text{s}}^{\text{e}}, P_{0}$ and fraction of signal time $\theta$,  
we can rewrite the outage probability minimization problem (\ref{opt_p:tet}) as
\begin{equation} \label{opt_delta:2}
\begin{aligned}
& \underset{\delta}{\text{minimize}}
& & \begin{cases}\vspace{0.3cm}
c_{11} \left(\frac{\delta}{1+\delta}\right)^\beta + c_{12} \left(\frac{1}{1+\delta}\right)^\beta  &\text{ $ \frac{P_{\text{proc}}+P_{\text{cons}}}{P_{\text{s}}^{\text{e}}(1-\theta)} \geq \frac{\gamma}{P_{\text{s}}^{\text{i}}\frac{\theta}{2}}$}\\ 
c_{21} \left(\frac{\delta}{1+\delta}\right)^\beta + c_{22} \left(\frac{1}{1+\delta}\right)^\beta &\text{ $\frac{P_{\text{proc}}+P_{\text{cons}}}{P_{\text{s}}^{\text{e}}(1-\theta)} \leq \frac{\gamma}{P_{\text{s}}^{\text{i}}\frac{\theta}{2}}$}
\end{cases} \\
& \text{subject to}
& & \delta\geq  \begin{cases}
\frac{\sqrt[\beta]{A}}{1-\sqrt[\beta]{A}}  &\text{  $A>0$}\\
0 &\text{ $A<0$}
\end{cases} ,
\end{aligned}
\end{equation}
where $c_{12}=c_{22}=\frac{\gamma 2^\beta}{P_{\text{r}}^{\text{i}}}$, $c_{11}=\frac{2^\beta(P_{\text{cons}}+P_{\text{proc}})\frac{\theta}{2}}{P_{\text{s}}^{\text{e}}(1-\theta)}$, $c_{21}=\frac{2^\beta \gamma}{P_{\text{s}}^{\text{i}}}$.\\
Both objective functions in (\ref{opt_delta:2}) follow the same form 
\begin{equation}
\begin{aligned}
f_{\text{i}}(\delta)&=c_{\text{i}1}\left(\frac{\delta}{1+\delta}\right)^\beta +c_{\text{i}2} \left(\frac{1}{1+\delta}\right)^\beta, i=1,2.
\end{aligned}
\end{equation}
Thus, from $\frac{\text{d}f_{\text{i}}(\delta)}{\text{d}\delta}= \beta \frac{c_{\text{i}1}\delta^{\beta-1}- c_{\text{i}2} }{(1+\delta)^{\beta+1}},$ the optimal relay position is determined by
\begin{equation} \label{eq:optimal_delta}
\delta^{\star} =  \begin{cases}
\max \left\lbrace\sqrt[\beta-1]{\frac{c_{\text{i}2}}{c_{\text{i}1}}},\frac{\sqrt[\beta]{A}}{1-\sqrt[\beta]{A}}\right\rbrace  &\text{ $A>0$}\\
\sqrt[\beta-1]{\frac{c_{\text{i}2}}{c_{\text{i}1}}} &\text{ $A<0$}\\  
\end{cases} .
\end{equation} 
Therefore, depending on the parameter settings, (\ref{eq:optimal_delta}) can be used to optimize $\delta$ in both branches of (\ref{opt_delta:2}). It is noteworthy that if $A>0$ and $\sqrt[\beta-1]{\frac{c_{\text{i}2}}{c_{\text{i}1}}}\leq \frac{\sqrt[\beta]{A}}{1-\sqrt[\beta]{A}}$, then the relay's optimal position is independent of the relay's transmission power. Besides, if $A\leq 0$ and the condition of the lower branch in (\ref{opt_delta:2}) holds, the optimal relay's position is independent of $\theta$.
The optimal position of the relay is further studied in Fig. \ref{fig:opt_pos_1} (Section \ref{nemodara}).
\subsection{Dynamic Time Sharing} \label{subsec:csi}
In Sections III.A-C, we assumed no CSI available at the source, motivated by the fact that the relay has no energy to estimate and feedback the CSI before the source energy transfer. In this subsection, we  relax this assumption and investigate the potential gains that the relay system can benefit from CSI feedback. Particularly, we consider the scheme where in each packet transmission the source energy transfer signal length $\theta$ is adaptively updated based on the instantaneous source-relay channel realization. Note that, as opposed to the fixed time sharing approach, the dynamic time sharing is based on the assumption that there is perfect CSI available at the source. This is an appropriate assumption in quasi-static conditions where the channel remains constant during multiple packet transmissions, so that the channel estimation delay/energy consumption can be ignored in the analysis.\par 
 In this protocol, the relay harvests as much energy such that it is active in the rest of packet transmission. In this case, the energy transfer period which is set adaptively by transmitter, is given by
\begin{equation} \label{eq:theta_dynam}
\begin{aligned}
&g_{\text{sr}}P_{\text{s}}^{\text{e}}(1-\theta)=P_{\text{cons}}\frac{\theta}{2}+E_{\text{proc}}
       \Rightarrow \theta=\frac{g_{\text{sr}}P_{\text{s}}^{\text{e}}-E_{\text{proc}}}{g_{\text{sr}}P_{\text{s}}^{\text{e}}+\frac{P_{\text{cons}}}{2}}.
\end{aligned}
\end{equation}
From (\ref{eq:theta_dynam}), it can be inferred that $\theta \leq 1$. Also, if $g_{\text{sr}}P_{\text{s}}^{\text{e}}\leq E_{\text{proc}}$ the relay is in the circuit outage. The outage probability for this scheme is given by  
\begin{equation}\label{eq:outage_dyna}
\small
\begin{aligned}
\mathrm{Pr}\left(\text{outage}\right)&=1-\int\limits_0^\infty \int\limits_0^\infty \mathrm{Pr}\left(xP_{\text{s}}^{\text{e}}\geq E_{\text{proc}},xP_{\text{s}}^{\text{i}}\geq \gamma,yP_{\text{r}}^{\text{i}}\geq \gamma\bigg|g_{\text{sr}}=x,g_{\text{rd}}=y \right)f_{g_{\text{sr}}}(x)f_{g_{\text{rd}}}(y)\text{d}x\text{d}y\\
&=1-\int\limits_{\Delta}^{\infty}\exp\left(-\frac{1}{\lambda_{rd}}\frac{\exp\left(2R\left(\frac{xP_{\text{s}}^{\text{e}}+\frac{P_{\text{cons}}}{2}}{xP_{\text{s}}^{\text{e}}-E_{\text{proc}}}\right)\right)-1}{P_{\text{r}}^{\text{i}}}\right)\frac{1}{\lambda_{\text{sr}}}\exp\left(-\frac{x}{\lambda_{sr}}\right)\text{d}x,
\end{aligned}
\end{equation}
where $\Delta=\max\left( \frac{E_{\text{proc}}}{P_{\text{s}}^{\text{e}}}, \xi \right)$ with $\xi$ defined as
\begin{equation}
\xi=\arg_{g_{\text{sr}}} \left\lbrace g_{\text{sr}}P_{\text{s}}^{\text{i}}-\exp\left(2R\left(\frac{g_{\text{sr}}P_{\text{s}}^{\text{e}}+\frac{P_{\text{cons}}}{2}}{g_{\text{sr}}P_{\text{s}}^{\text{e}}-E_{\text{proc}}}\right)\right)+1=0 \right\rbrace.
\end{equation}
The integration (\ref{eq:outage_dyna}) does not have a closed-form expression in general. However, considering the high-SNR regime, the outage probability in (\ref{eq:outage_dyna}) is obtained as
\begin{equation}\label{eq:alakie_hsnr}
 \mathrm{Pr}\left(\text{outage}\right) \simeq 1 - \exp\left(-\frac{\exp\left(2R\right)-1}{\lambda_{\text{rd}} P_{\text{r}}^{\text{i}}}\right).
\end{equation}
It can be inferred from (\ref{eq:shirr}) and (\ref{eq:alakie_hsnr}) that the high-SNR outage probabilities of dynamic time sharing and fixed time sharing provided that the relay utilizes optimal time sharing in (\ref{eq:opt_thettttt}) are the same since, at high SNRs, the optimal time sharing converges to 1.  However, at low/moderate SNRs dynamic time sharing outperforms fixed time sharing, in terms of the outage probability, as expected. The performance comparison of the fixed and dynamic time sharing protocols are further studied in Fig. \ref{fig:out_dyn_fix}.
\vspace{-3mm}
\section{On the effect of relay's power adaptation}\label{sec:adp}
Here, we assume that the relay can adaptively update the transmission power in the third phase and uses all available energy to forward the source message with maximum power. Thus, the relay transmission power is obtained by 
\begin{equation} \label{eq:trans_pow}
\begin{aligned}
&(P_{\text{active}}+\nu P_{\text{r}}^{\text{i}})\frac{\theta}{2}=g_{\text{sr}}P_{\text{s}}^{\text{e}}(1-\theta)-P_{\text{proc}}\frac{\theta}{2}
 \Rightarrow  P_{\text{r}}^{\text{i}}=\frac{g_{\text{sr}}P_{\text{s}}^{\text{e}}\left(1-\theta\right)-\left(P_{\text{proc}}+P_{\text{active}}\right)\frac{\theta}{2}}{\nu \frac{\theta}{2}},
\end{aligned}
\end{equation}
and the outage probability is rephrased as
\begin{equation} \label{eq:basic_opor_outage}
\begin{aligned}
\mathrm{Pr}\left(\text{outage}\right)&=1-\mathrm{Pr}\bigg( g_{\text{sr}} \geq \frac{\left(P_{\text{proc}}+P_{\text{active}}\right)\theta}{2\left(1-\theta\right)P_{\text{s}}^{\text{e}}}, g_{\text{sr}} \geq \frac{\gamma}{P_{\text{s}}^{\text{i}}} \\
& \hspace{2cm} ,g_{\text{rd}}\left(g_{\text{sr}}-\frac{\left(P_{\text{proc}}+P_{\text{active}}\right)\theta}{2\left(1-\theta\right)P_{\text{s}}^{\text{e}}}\right)\geq \frac{\gamma \nu \theta}{2\left(1-\theta\right)P_{\text{s}}^{\text{e}}} \bigg)\\
&=1-\mathrm{Pr}\left(g_{\text{sr}} \geq \frac{\alpha_1}{P_{\text{s}}^{\text{e}}}, g_{\text{sr}} \geq \frac{\alpha_2}{P_{\text{s}}^{\text{i}}} , g_{\text{rd}}\left(g_{\text{sr}}-\frac{\alpha_1}{P_{\text{s}}^{\text{e}}}\right)\geq \frac{\alpha_3}{P_{\text{s}}^{\text{e}}} \right),
\end{aligned}
\end{equation}
where in the last equality,
\begin{equation}\label{eq:deff}
\begin{aligned}
\alpha_1=\frac{\left(P_{\text{proc}}+P_{\text{active}}\right)\theta}{2(1-\theta)},
\alpha_2=\gamma,
\alpha_3=\frac{\gamma\nu \theta }{2(1-\theta)},
\end{aligned}
\end{equation}
  are constants determined by the system parameters.
Moreover, we have
\begin{equation} \label{eq:nRc_be_1}
\begin{small}
\begin{aligned}
\mathrm{Pr}\left(\text{Outage}\right)&=1-\int\limits_{0}^{\infty}\int\limits_{0}^{\infty}\mathrm{Pr}\left(x \geq \frac{\alpha_2}{P_{\text{s}}^{\text{i}}},y\left(x-\frac{\alpha_1}{P_{\text{s}}^{\text{e}}}\right)\geq \frac{\alpha_3}{P_{\text{s}}^{\text{e}}},x \geq \frac{\alpha_1}{P_{\text{s}}^{\text{e}}} \bigg|g_{\text{sr}}=x,g_{\text{rd}}=y\right)f_{g_{\text{sr}}}\left(x\right)f_{g_{\text{rd}}}\left(y\right)\text{d}x\text{d}y\\
&=1-
\frac{\alpha_3}{\lambda_{\text{sr}}\lambda_{\text{rd}}P_{\text{s}}^{{\text{e}}}}\int \limits_{\mathcal{Q}}^{\infty}\exp\left(-\frac{1}{u}\right)\exp\left(-\frac{\alpha_3u+\lambda_{\text{rd}}\alpha_1}{\lambda_{\text{sr}}\lambda_{\text{rd}}P_{\text{s}}^{{\text{e}}}}\right) \text{d}u,
\end{aligned}
\end{small}
\end{equation}
where $\mathcal{Q}=\max\left\lbrace 0,\lambda_{\text{rd}} \left(\frac{P_{\text{s}}^{\text{e}}\alpha_2}{P_{\text{s}}^{\text{i}}\alpha_3}-\frac{\alpha_1}{\alpha_3}\right)\right\rbrace$. Here, the last equality comes from the variable transformation $u=\lambda_{\text{rd}}\frac{P_{\text{s}}^{\text{e}}x-\alpha_1}{\alpha_3}$. In this way, the integral in (\ref{eq:nRc_be_1}) needs to be calculated in the following two cases: $\frac{\alpha_2}{P_{\text{s}}^{\text{i}}}\geq \frac{\alpha_1}{P_{\text{s}}^{\text{e}}}$ and $\frac{\alpha_2}{P_{\text{s}}^{\text{i}}}\leq \frac{\alpha_1}{P_{\text{s}}^{\text{e}}}$ .
Considering $\frac{\alpha_2}{P_{\text{s}}^{\text{i}}}\leq \frac{\alpha_1}{P_{\text{s}}^{\text{e}}}$, we have
\begin{equation} \label{eq:outage_1_adaptive}
\begin{aligned}
\mathrm{Pr}\left(\text{outage}\right)&=1-\frac{\alpha_3}{\lambda_{\text{sr}}\lambda_{\text{rd}}P_{\text{s}}^{{\text{e}}}}\exp\left(-\frac{\alpha_1}{\lambda_{\text{sr}}P_{\text{s}}^{{\text{e}}}}\right)\int\limits_{0}^{\infty}\exp\left(-\frac{\alpha_3u}{\lambda_{\text{sr}}\lambda_{\text{rd}}P_{\text{s}}^{{\text{e}}}}\right)\text{d}u\\
&= 1-\sqrt{\frac{4\alpha_3}{\lambda_{\text{sr}}\lambda_{\text{rd}}P_{\text{s}}^{{\text{e}}}}}\exp\left(-\frac{\alpha_1}{\lambda_{\text{sr}}P_{\text{s}}^{{\text{e}}}}\right)\text{K}_1\left(\sqrt{\frac{4\alpha_3}{\lambda_{\text{sr}}\lambda_{\text{rd}}P_{\text{s}}^{{\text{e}}}}}\right),
\end{aligned}
\end{equation}
where the last equality is obtained by the definition of the modified Bessel function of second kind \cite[Eq. 3.324.1]{tabebesel}. On the other hand, if $\frac{\alpha_2}{P_{\text{s}}^{\text{i}}}\geq \frac{\alpha_1}{P_{\text{s}}^{\text{e}}}$, the outage probability is given by
\begin{equation} \label{eq:inttt_badeeee}
\begin{aligned}
\small
\mathrm{Pr}(\text{outage})&=1-\frac{\alpha_3}{\lambda_{\text{sr}}\lambda_{\text{rd}}P_{\text{s}}^{\text{e}}}\exp\left(-\frac{\alpha_1}{\lambda_{\text{sr}}P_{\text{s}}^{\text{e}}}\right)\int \displaylimits_{\Lambda}^{\infty} \exp\left(-\frac{1}{u}\right)\exp\left(-\frac{\alpha_3u}{\lambda_{\text{sr}}\lambda_{\text{rd}}P_{\text{s}}^{{\text{e}}}}\right)\text{d}u\\
& \simeq 1-\frac{\alpha_3}{\lambda_{\text{sr}}\lambda_{\text{rd}}P_{\text{s}}^{\text{e}}}\exp\left(-\frac{\alpha_1}{\lambda_{\text{sr}}P_{\text{s}}^{\text{e}}}\right)\int\limits_{\Lambda}^{\infty}\sum\limits_{m=0}^{M}\frac{(-1)^m}{m!}\frac{1}{u^m}\exp\left(-\frac{\alpha_3u}{\lambda_{\text{sr}}\lambda_{\text{rd}}P_{\text{s}}^{{\text{e}}}}\right)\text{d}u\\
&= 1-\frac{\alpha_3}{\Lambda\lambda_{\text{sr}}\lambda_{\text{rd}}P_{\text{s}}^{\text{e}}}\exp\left(-\frac{\alpha_1}{\lambda_{\text{sr}}P_{\text{s}}^{\text{e}}}\right)\sum\limits_{m=0}^{M}\left(-\Lambda\right)^{-m} \text{E}_m\left(-\frac{\alpha_3 \Lambda}{\lambda_{\text{sr}}\lambda_{\text{rd}}P_{\text{s}}^{\text{e}}}\right),
\end{aligned}
\end{equation}
for all $M \geq 1$, where
$
\Lambda=\lambda_{\text{rd}} \left(\frac{P_{\text{s}}^{\text{e}}\alpha_2}{P_{\text{s}}^{\text{i}}\alpha_3}-\frac{\alpha_1}{\alpha_3}\right)$, $\alpha_1,\alpha_2,\alpha_3$ are defined in (\ref{eq:deff}), the approximation comes from the Taylor expansion of the exponential function and the last equality is obtained by the definition of generalized exponential integral\cite[Eq. 8.211.1]{tabebesel}.
The overall outage probability for the power-adaptive relay from (\ref{eq:outage_1_adaptive}) and (\ref{eq:inttt_badeeee})  is given by
\begin{equation} \label{eq:ove_out_pro_adap}
\begin{aligned}
\mathrm{Pr}(\text{outage})&=\begin{cases}\vspace{0.15cm}
1-\sqrt{\frac{4\alpha_3}{\lambda_{\text{sr}}\lambda_{\text{rd}}P_{\text{s}}^{{\text{e}}}}}\exp\left(-\frac{\alpha_1}{\lambda_{\text{sr}}P_{\text{s}}^{{\text{e}}}}\right)\text{K}_1\left(\sqrt{\frac{4\alpha_3}{\lambda_{\text{sr}}\lambda_{\text{rd}}P_{\text{s}}^{{\text{e}}}}}\right) &\text{ $\frac{\alpha_2}{P_{\text{s}}^{\text{i}}}\leq \frac{\alpha_1}{P_{\text{s}}^{\text{e}}}$}\\ 
1-\frac{\alpha_3}{\mathcal{V}\lambda_{\text{sr}}\lambda_{\text{rd}}P_{\text{s}}^{\text{e}}}\exp\left(-\frac{\alpha_1}{\lambda_{\text{sr}}P_{\text{s}}^{\text{e}}}\right)\sum\limits_{m=0}^{M}\left(-\mathcal{V}\right)^{-m} \text{E}_m\left(-\frac{\alpha_3 \mathcal{V}}{\lambda_{\text{sr}}\lambda_{\text{rd}}P_{\text{s}}^{\text{e}}}\right)&\text{ $\frac{\alpha_2}{P_{\text{s}}^{\text{i}}}\geq \frac{\alpha_1}{P_{\text{s}}^{\text{e}}}$}
\end{cases}.
\end{aligned}
\end{equation}
While the detailed discussion is omitted due to space limitation, we can show that
 the adaptive power allocation for the relay has the diversity order of one.
\section{Performance Analysis in Multi-Relay Networks}\label{sec:multi}
In this section, we extend our analysis to the multi-relay networks where $N$ relays are employed to assist the information transmission between the source and the destination. We analyze the system performance for the opportunistic relaying \cite{opor_relay} with perfect and partial CSI available at the relays. Also, for brevity, we evaluate the performance for the adaptive power allocation at the relays. Note that the subsequent analysis can be readily extended to the case with non-adaptive power allocation at the relays. \\
In the opportunistic relaying with perfect CSI, the selection is performed considering the source-relay and the relay-destination CSI. That is, the  connecting relay is selected according to
\begin{equation} \label{eq:opor_dual_hop}
i^{*}=\underset{1 \leq i \leq N}{\text{argmax}}\{\min\{g_{\text{sr}_i},g_{\text{r}_i\text{d}}\}\}.
\end{equation}
On the other hand, if the relays are able to acquire either the source-destination or relay-destination CSI, which we call as partial CSI throughout the paper,
the connecting relay is selected based on one of the links' CSI. Particularly, when the source-relay CSI is considered, the connecting relay is selected as
\begin{equation} \label{eq:opor_hop_1}
i^{*}=\underset{1 \leq i \leq N}{\text{argmax}}\{ g_{\text{sr}_i}\}.
\end{equation}  
Similarly, when the CSI of the relay-destination channel is available, the connecting relay is selected as 
\begin{equation}\label{eq:opor_hop_2}
i^{*}=\underset{1 \leq i \leq N}{\text{argmax}}\{ g_{\text{r}_i\text{d}}\}. 
\end{equation}
It is worth noting that the best relay selection can be implemented based on the local measurements of the instantaneous channel conditions at the relays and hence the selection scheme of (\ref{eq:opor_dual_hop}), (\ref{eq:opor_hop_1}) and (\ref{eq:opor_hop_2}) can be implemented in the distributed manner\cite{distributed1}.\\
 In the following subsections, the performance of each selection criterion is investigated.
\subsection{Performance Analysis of Opportunistic Relaying with Perfect CSI}
In this subsection, we assume that the relays can obtain the channel power gains, i.e., $g_{\text{sr}}$ and $g_{\text{rd}}$, before the entire transmission. Then, the selected relay, based on the criteria in (\ref{eq:opor_dual_hop}), employs the protocol as presented in Section \ref{sec:adp} to assist the communication. In Theorem \ref{thr:opor_full} and Corollary \ref{corr:opor_full}, we investigate the outage probability of this scheme and its high-SNR approximation, respectively.
\begin{theorem} \label{thr:opor_full}
The outage probability of the opportunistic relaying with perfect CSI is given by
\end{theorem}
\vspace{-20pt}
\begin{equation}\label{eq:kokoli}
\begin{aligned}
&\mathrm{Pr}\left(\text{outage}\right)^{\text{Perfect-CSI}}=\\
& \begin{cases}
\left(1-\exp\left(-\left(\frac{1}{\lambda_{\text{sr}}}+\frac{1}{\lambda_{\text{rd}}}\right)\widetilde{\alpha}\right)\right)^N-\left[ N\sum\limits_{k=0}^{N-1} {N-1 \choose k}(-1)^k\left(\frac{1}{\lambda_{\text{sr}}}\int\limits_{\frac{\alpha_1}{P_{\text{s}}^{\text{e}}}}^{\widetilde{\alpha}}J_k\left(x\right)\text{d}x +\frac{1}{\lambda_{\text{rd}}}\int\limits_{0}^{\widetilde{\alpha}}U_k\left(x\right)\text{d}x\right)\right] \\
\hspace{10cm} \text{ if $\frac{\alpha_1}{P_{\text{s}}^{\text{e}}}\geq \frac{\alpha_2}{P_{\text{s}}^{\text{i}}}$}\\
\left(1-\exp\left(-\left(\frac{1}{\lambda_{\text{sr}}}+\frac{1}{\lambda_{\text{rd}}}\right)\frac{\alpha_2}{P_{\text{s}}^{\text{i}}}\right)\right)^N \text{ if $\frac{\alpha_1}{P_{\text{s}}^{\text{e}}}<\frac{\alpha_2}{P_{\text{s}}^{\text{i}}}$ $\&$ $\frac{\alpha_2}{P_{\text{s}}^{\text{i}}}\geq \widetilde{\alpha} $}\\
 \left(1-\exp\left(-\left(\frac{1}{\lambda_{\text{sr}}}+\frac{1}{\lambda_{\text{rd}}}\right)\widetilde{\alpha}\right)\right)^N-\left[N\sum\limits_{k=0}^{N-1} {N-1 \choose k}(-1)^k\left(\frac{1}{\lambda_{\text{sr}}}\int\limits_{\frac{\alpha_2}{P_{\text{s}}^{\text{i}}}}^{\widetilde{\alpha}}J_k\left(x\right)\text{d}x+\frac{1}{\lambda_{\text{rd}}}\int\limits_{\frac{\alpha_2}{P_{\text{s}}^{\text{i}}}}^{\widetilde{\alpha}}U_k\left(x\right)\text{d}x\right)\right]\\
 \hspace{10cm}  \text{ if $\frac{\alpha_1}{P_{\text{s}}^{\text{e}}}<\frac{\alpha_2}{P_{\text{s}}^{\text{i}}}$ $\&$ $\frac{\alpha_2}{P_{\text{s}}^{\text{i}}}< \widetilde{\alpha} $}
\end{cases}, 
\end{aligned}
\end{equation} 
where $J_{k}\left(x\right)\triangleq \exp\left(-\frac{ \alpha_3}{\lambda_{\text{rd}}\left(P_{\text{s}}^{\text{e}}x-\alpha_1\right)}-\frac{(k+1)x}{\lambda_{\text{sr}}}-\frac{kx}{\lambda_{\text{rd}}}\right)$, $U_{k}\left(x\right) \triangleq \exp\left(-\frac{1}{\lambda_{\text{sr}}} \left(\frac{\alpha_3}{P_{\text{s}}^{\text{e}}x}+\frac{\alpha_1}{P_{\text{s}}^{\text{e}}}\right)-\frac{kx}{\lambda_{\text{sr}}} -\frac{(k+1)x}{\lambda_{\text{rd}}} \right)$ and $\widetilde{\alpha}\triangleq\frac{\alpha_1+\sqrt{\alpha_1^2+4P_{\text{s}}^{\text{e}}\alpha_3}}{2P_{\text{s}}^{\text{e}}}$. Also, $\alpha_1,\alpha_2$, and $\alpha_3$ are defined in (\ref{eq:deff}).
\begin{proof}
See Appendix C.
\end{proof}
\begin{coro} \label{corr:opor_full}
Let $P_{\text{s}}^{\text{i}}=P_{\text{s}}^{\text{e}}=P$. At the high SNRs, the outage probability in (\ref{eq:kokoli}) is approximated by
\begin{equation} \label{eq:diverr_opor}
\mathrm{Pr}\left(\text{outage}\right)^{\text{Perfect-CSI}}\simeq \Omega_N \left(\frac{\gamma\nu \theta }{2(1-\theta)}\right)^{\frac{N+1}{2}} P^{-\left(\frac{N+1}{2}\right)},
\end{equation}
where
\begin{equation}\label{eq:muuu}
\begin{aligned}
\small
&\Omega_N \triangleq \left[\frac{S_{N+1}(-1)^{N+1}\left(\lambda_{\text{sr}}+\lambda_{\text{rd}}\right)^{N+1}}{(N+1)!\left(\lambda_{\text{sr}}\lambda_{\text{rd}}\right)^{N+1}}-\frac{N(-1)^N\left(\lambda_{\text{sr}}^{N+1}+\lambda_{\text{rd}}^{N+1}\right)}{\left(N+1\right)!\left(\lambda_{\text{sr}}\lambda_{\text{rd}}\right)^{N+1}}\left(N\Phi_{N-1}+\Phi_N\right)+\frac{2N\left(\lambda_{\text{sr}}+\lambda_{\text{rd}}\right)^{N-1}}{(N-1)\left(\lambda_{\text{sr}}\lambda_{\text{rd}}\right)^{N}}\right],  \\
&S_{N+1}\triangleq \sum\limits_{k=0}^{N}{N \choose k}(-1)^k k^{N+1},\Phi_{l}\triangleq \left(\frac{\lambda_{\text{sr}}+\lambda_{\text{rd}}}{\lambda_{\text{rd}}}\right)^l\left[\sum\limits_{k=0}^{N-1}{N-1 \choose k}(-1)^kk^l\right].
\end{aligned}
\end{equation}
\begin{proof}
See Appendix D.
\end{proof}
\end{coro}
From (\ref{eq:diverr_opor}), it can be inferred that the diversity order of opportunistic relaying with perfect CSI is $\frac{N+1}{2}$. This is different from the cases with fixed-energy-supply relay networks, where the opportunistic relaying scheme achieves the maximal diversity gain, i.e., $N$\cite{opor_relay}. Also, our simulation results in Fig. \ref{fig:mult_rel_2} indicate that, the high-SNR approximation of (\ref{eq:diverr_opor}) is tight for the broad range of SNRs. From (\ref{eq:diverr_opor}), it is worth noting that, since $\alpha_3$ is linearly proportional to the PA inefficiency (\ref{eq:deff}), i.e. $\nu$, the outage probability increases in $\nu$ with power of $\frac{N+1}{2}$. Hence, the inefficiency of the PA remarkably affects the outage performance. 
\subsection{Performance Analysis of Opportunistic Relaying with Partial CSI}
Since acquiring the perfect CSI of the both channels incurs extra overhead, in this subsection, we analyze the impact of the availability of either the source-relay or the relay-destination CSI on the system performance. 
\begin{theorem} \label{thr:opor_partial}
The outage probability of opportunistic relaying with knowledge of the source-relay CSI and the relay-destination CSI is given by
\end{theorem}
\vspace{-15pt}
\begin{align}
\small
&\mathrm{Pr}\left(\text{outage}\right)^{\text{SR-CSI}}=\begin{cases} \vspace{0.3cm}
1-N \sum\limits_{i=0}^{N-1}(-1)^i {N-1 \choose i}  \exp\left(-\frac{(i+1)\alpha_1}{\lambda_{\text{sr}} P_{\text{s}}^{\text{e}}}\right)\sqrt{\frac{4 \alpha_3 }{P_{\text{s}}^{\text{e}} \lambda_{\text{sr}} \lambda_{\text{rd}}(i+1)}}\text{K}_1\left(\frac{4\alpha_3 (i+1) }{P_{\text{s}}^{\text{e}}\lambda_{\text{sr}} \lambda_{\text{rd}}}\right) \\ 
\hspace{9cm} \text{ if $\frac{\alpha_1}{P_{\text{s}}^{\text{e}}} \geq \frac{\alpha_2}{P_{\text{s}}^{\text{i}}}$}  \\
1-N\sum\limits_{i=0}^{N-1}\frac{(-1)^i}{\lambda_{\text{sr}}} {N-1 \choose i} \exp\left(-\frac{(i+1)\alpha_1}{\lambda_{\text{sr}} P_{\text{s}}^{\text{e}}}\right)  \int\limits_{\frac{\alpha_2}{P_{\text{s}}^{\text{i}}}-\frac{\alpha_1}{P_{\text{s}}^{\text{e}}}}^{\infty} \exp\left(-\frac{\lambda_2 \alpha_3}{\lambda_{\text{rd}}P_{\text{s}}^{\text{e}}x}-\frac{(i+1)x}{\lambda_{\text{sr}}}\right)\text{d}x \\
\hspace{9cm} \text{ if $\frac{\alpha_2}{P_{\text{s}}^{\text{i}}} \geq \frac{\alpha_1}{P_{\text{s}}^{\text{e}}}$}\\  
\end{cases} \label{eq:hop1_out}\\
\vspace{-10pt}
\text{and} \nonumber\\
\vspace{-10pt}
&\mathrm{Pr}\left(\text{outage}\right)^{\text{RD-CSI}}=\begin{cases} \vspace{0.3cm}
1+  \exp\left(-\frac{\alpha_1}{\lambda_{\text{sr}} P_{\text{s}}^{\text{e}}}\right) \sum\limits_{i=1}^{N}(-1)^i {N \choose i} \sqrt{\frac{4 \alpha_3 i }{\lambda_{\text{sr}} \lambda_{\text{rd}} P_{\text{s}}^{\text{e}} }}\text{K}_1\left(\frac{4\alpha_3 i }{\lambda_{\text{sr}} \lambda_{\text{rd}}P_{\text{s}}^{\text{e}}}\right) \quad \quad
\text{ if $\frac{\alpha_1}{P_{\text{s}}^{\text{e}}} \geq \frac{\alpha_2}{P_{\text{s}}^{\text{i}}}$}\\
1+\exp\left(-\frac{\alpha_2}{\lambda_{\text{sr}} P_{\text{s}}^{\text{i}}}\right)\sum\limits_{i=1}^{N} {N \choose i} \frac{(-1)^i }{\lambda_{\text{sr}}}\int\limits_{0}^{\infty}\exp \left(-\frac{i\frac{\alpha_3}{P_{\text{s}}^{\text{e}}}}{\lambda_{\text{rd}}\left(x+\left(\frac{\alpha_2}{P_{\text{s}}^{\text{i}}}-\frac{\alpha_1}{P_{\text{s}}^{\text{e}}}\right)\right)}-\frac{x}{\lambda_{\text{sr}}}\right)\text{d}x \\
\hspace{9cm} \text{ if $\frac{\alpha_2}{P_{\text{s}}^{\text{i}}} \geq \frac{\alpha_1}{P_{\text{s}}^{\text{e}}}$} \\  
\end{cases} ,\label{eq:hop2_out}
\end{align}
respectively. In (\ref{eq:hop2_out}), $\text{K}_{1}(.)$ denotes the modified Bessel function of the second kind, $\alpha_1,\alpha_2$, and $\alpha_3$ are defined in (\ref{eq:deff}).
\begin{proof}
The proof is similar to the proof of Theorem \ref{thr:opor_full} and omitted for the sake of brevity. 
\end{proof}
\begin{coro} \label{corr:opor_partial}
Let $P_{\text{s}}^{\text{i}}=P_{\text{s}}^{\text{e}}=P$. At high SNRs, the outage for the partial CSI is obtained as
\end{coro}
\vspace{-30pt}
\begin{align}
&\mathrm{Pr}\left(\text{outage}\right)^{\text{SR-CSI}}\simeq \frac{ \theta }{2  \left(1-\theta\right) \lambda_{\text{sr}}  P} \ \frac{ \gamma \nu}{\lambda_{\text{rd}}}\Xi_N . \label{eq:hop1_high_snr}\\
&\mathrm{Pr}\left(\text{outage}\right)^{\text{RD-CSI}}\simeq \begin{cases} \vspace{0.2cm}
\frac{\theta}{2\left(1-\theta\right)\lambda_{\text{sr}} P}\left(\frac{\gamma \nu }{\lambda_{\text{rd}}} \ \Xi_N + \left(P_{\text{proc}}+P_{\text{active}}\right) \right) &\text{ $\frac{P_{\text{proc}}+P_{\text{cons}}}{(1-\theta)} \geq \frac{2 \gamma}{\theta}$}\\
\frac{\theta}{2\left(1-\theta\right)\lambda_{\text{sr}} P}\left(\frac{\gamma \nu }{\lambda_{\text{rd}}} \ \Xi_N + \gamma \right)   &\text{ $\frac{P_{\text{proc}}+P_{\text{cons}}}{(1-\theta)} \leq \frac{2 \gamma}{\theta}$} \\  
\end{cases}, \label{eq:hop2_high_snr}
\end{align}
where $ \Xi_N=N\sum\limits_{i=0}^{N-1} {N-1 \choose i} (-1)^{(i+1)} \log\left(i+1\right)$. 
\begin{proof}
See Appendix E.
\end{proof}
Equations (\ref{eq:hop1_high_snr}) and (\ref{eq:hop2_high_snr}) imply that opportunistic relaying with partial CSI has the diversity order of one. Moreover, at high SNRs, the relay selection based on the source-relay CSI outperforms the relay selection based on the relay-destination CSI. Our further results in Figs. \ref{fig:mult_rel_1} and \ref{fig:mult_rel_2} indicate that the superiority of relay selection criteria in (\ref{eq:opor_hop_1}) over (\ref{eq:opor_hop_2}) is valid for all SNRs. Finally, from (\ref{eq:hop1_high_snr}) and (\ref{eq:hop2_high_snr}), as the expected gain of the relay-destination channel decreases, the same performance is observed in both partial CSI models, as expected.
\section{Simulation Results} \label{nemodara}
In all figures, expect Fig. \ref{fig:opt_pos_1} which analyzes the optimal relay position, we consider fading channels with $\lambda_{\text{sr}}=\lambda_{\text{rd}}=1$ and, in harmony with \cite{paperb3,10,poor,13,encoop,100,tars}, we do not consider the large scale fading. The impact of large scale fading is studied in Fig. \ref{fig:opt_pos_1}. Also, we set $P_{\text{proc}}=200$ mW $=23.01$ dBm, $P_{\text{active}}=800$ mW$=29.03$ dBm \cite{paperjin1,paper_AB}. The slope of the power consumption model in  (\ref{eq:power_cons_model}) is set to $\nu=2$ which is typical for class-AB amplifiers \cite{paperjin1,paper_AB}. Figures 1-6 present the results for the cases with single relays. The effect of multi-relays on the system performance is studied in Figs. 7-9. \par
\begin{figure}
\centering     
\includegraphics[width=0.95\columnwidth]{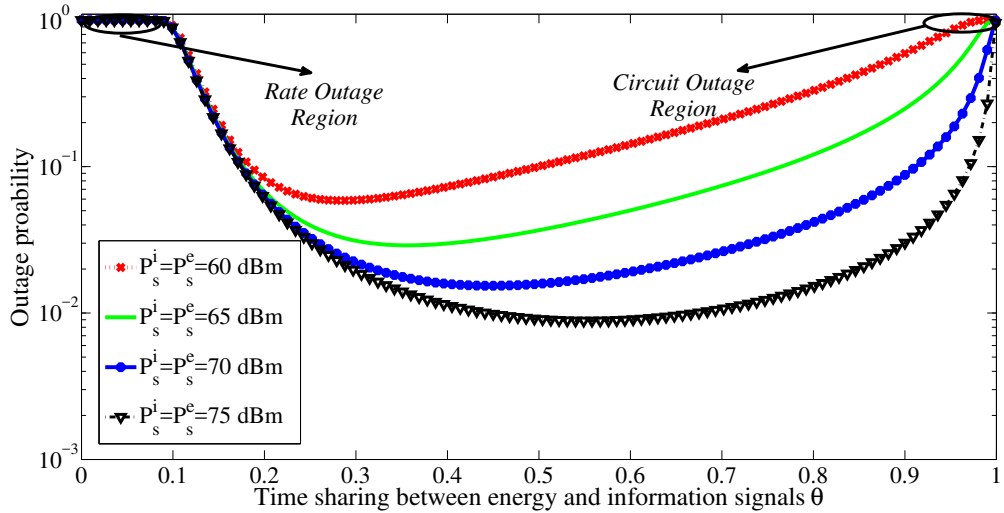}
\label{fig:theta_diff_pow}
\caption{Outage probability versus time sharing for different source transmission powers, $\lambda_{\text{sr}}=\lambda_{\text{rd}}=1$, $\nu=2$, $P_{\text{proc}}=23.01$ dBm, $P_{\text{active}}=29.03$ dBm, $P_{\text{r}}^{\text{i}}=55$ dBm. }
\label{fig:theta_diff_rate}
\end{figure}
\textit{On the Outage Probability of the Non-Adaptive Power Allocation Scheme:} Fig. \ref{fig:theta_diff_rate} shows the outage probabilities of the non-adaptive power allocation scheme, given in (\ref{eq:ov_out_f}), versus the time sharing between the length of energy and information signals, i.e., $\theta$. We have set the relay's transmission power to $55$ dBm, i.e., $P_{\text{r}}^{\text{i}}=55$ dBm. Fig. \ref{fig:theta_diff_rate} studies the outage probability for 4 pairs of $(P_{\text{s}}^{\text{i}},P_{\text{s}}^{\text{e}})$ by setting the codeword rate to $0.5$ bit per channel use (bpcu). Based on the segmentation of each communication block, the outage event for small $\theta$ is due to the rate outage event, since for $\theta \rightarrow 0$, a small amount of the block time is allocated for data transmission. On the other hand, for large values of $\theta$, the circuit outage event occurs since the energy harvesting period in each block is $1-\theta$. Hereby, we can define "circuit outage region" and "rate outage region" as indicated in Fig. \ref{fig:theta_diff_rate}. In the rate outage region and the circuit outage region, the outage probability is primarily dependent on the codeword rate and source transmission power, respectively (Fig. 1).\par
\begin{figure}
\begin{center}
\includegraphics[width=0.9\columnwidth]{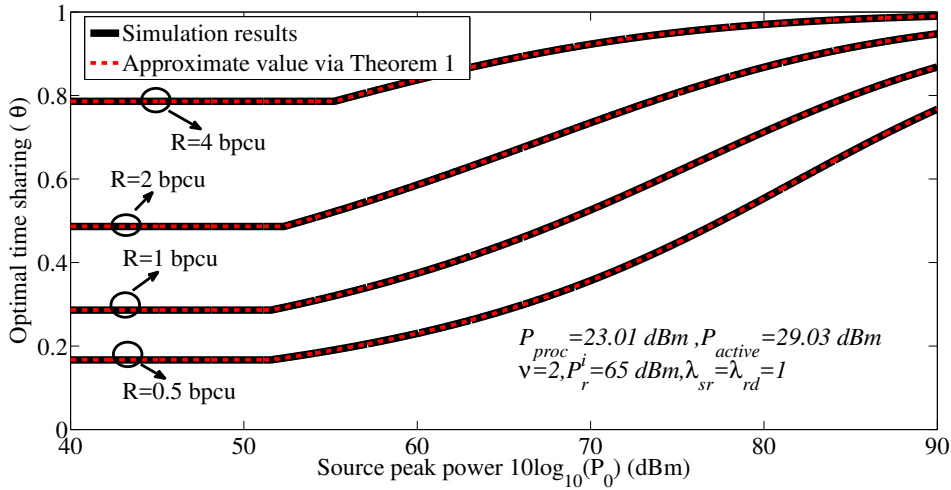}
\caption{Optimal time sharing versus source peak transmission power.} \label{fig:optimal_theta}
\end{center}
\end{figure}

\textit{Optimal Time Sharing:} Figure \ref{fig:optimal_theta} considers the optimal, in terms of (\ref{opt_p:tet}), time sharing for a predetermined source peak power which is given in Theorem \ref{thr:tet} and compares the results with the ones derived by exhaustive search. Here, the results are presented for $R=\{0.5,1,2,4\}$ bpcu. As the codeword rate increases, the optimal time sharing allocated for the information transmission increases since the transmission rate is proportional to $\theta$.
Optimal time sharing has two regions. In the first region, the optimal value of $\theta$ is independent of
the source transmission power, which is given in lower branch
of (\ref{eq:opt_thettttt}). In the second region, the optimal time sharing is an increasing function of the
source transmission power and tends towards one as the source
transmission power increases. The approximation technique of Theorem \ref{thr:tet} is very tight
for a broad range of source transmission powers/codeword rates. \par
\begin{figure}
\begin{center}
\includegraphics[width=0.9\columnwidth]{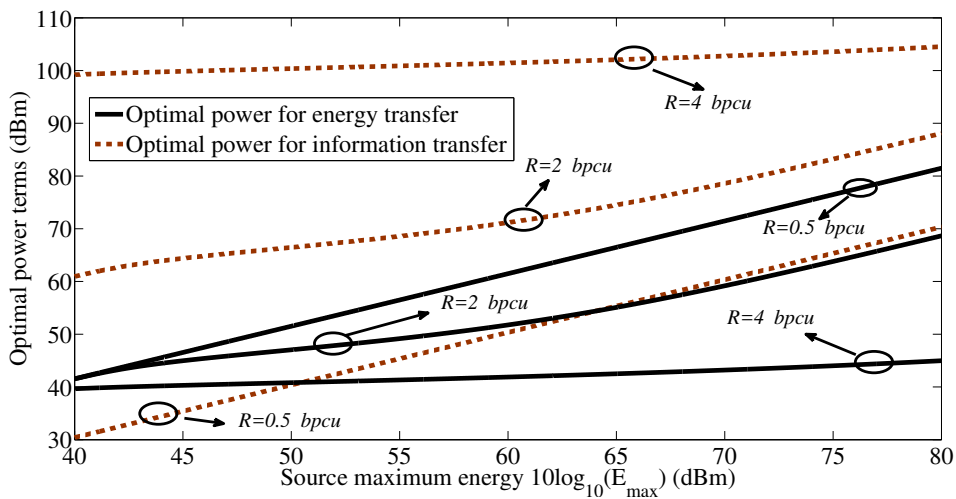}
\caption{Optimal power allocation versus the source energy for different codeword rates, $\lambda_{\text{sr}}=\lambda_{\text{rd}}=1$, $\nu=2$, $P_{\text{proc}}=23.01$ dBm, $P_{\text{active}}=29.03$ dBm, $P_{\text{r}}^{\text{i}}=55$ dBm.} \label{fig:opt_pow_1}
\end{center}
\end{figure}
\begin{figure}
\begin{center}
\includegraphics[width=0.9\columnwidth]{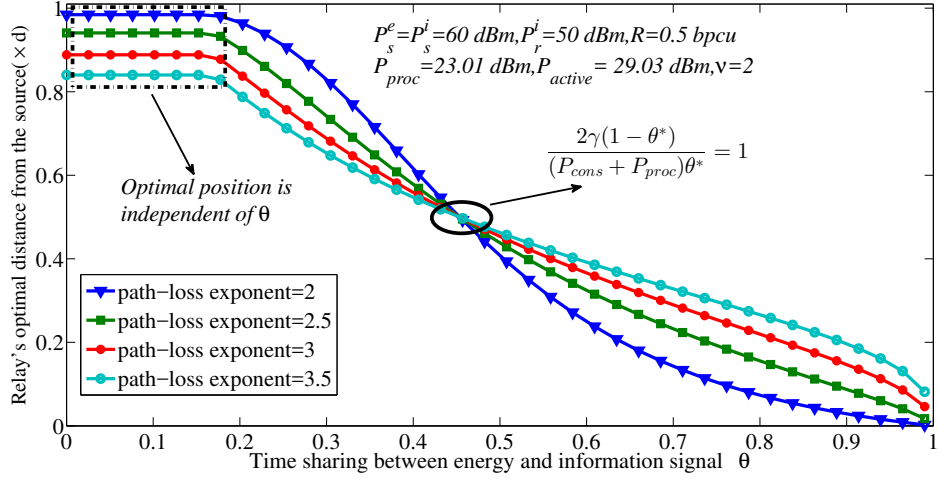}
\caption{Optimal relay position versus the source transmission power.} \label{fig:opt_pos_1}
\end{center}
\end{figure}

\textit{Optimal Power Allocation at the Source:} Figure \ref{fig:opt_pow_1} investigates optimal power allocation at the source versus the source maximum energy constraint, derived in Theorem 2, for different
codeword rates. The results are also double-checked using exhaustive search. Increasing the codeword rate, more power is allocated to the information signal, and optimal power of the energy signal decreases exponentially with the codeword rate. Also, in the log-log domain,  the optimal power term for the energy transfer increases linearly with the source maximum energy at high SNRs, as also stated in Eq. (\ref{eq:opt_p_hsnr}). \par
\textit{On the Optimal Relay Position:} Setting $P_{\text{s}}^{\text{i}}=P_{\text{s}}^{\text{e}}=70$ dBm, $P_{\text{r}}^{\text{i}}=50 $ dBm , $R=0.5$ bpcu, Fig. \ref{fig:opt_pos_1} shows the optimal relay's position versus the time sharing for the pathloss exponents of $\{2,2.5,3,3.5\}$. As seen in Fig. \ref{fig:opt_pos_1}, the optimal position is a decreasing function of $\theta$. Intuitively, since as $\theta$ increases, 
the circuit outage event is likely to occur, so, in the optimal case, the relay should be close to the source to maximize the harvested energy. Moreover, there exists a value for the time sharing for which the relay's optimal position is the same for all pathloss exponents. In that point, according to the parameter setting, we have $A\leq 0$ , so the first branch condition in (\ref{opt_delta:2}) holds and $\frac{2\gamma\left(1-\theta\right)}{(P_{\text{proc}}+P_{\text{cons}})\theta}=1$. Hence, the optimal relay position is $\frac{d}{2}$, i.e., mid-point of the source-destination (\ref{eq:optimal_delta}) for all the pathloss exponent.
Also, as seen in (\ref{eq:optimal_delta}), there exists a region for $\theta$ such that the optimal relay position is independent of $\theta$ (Fig. \ref{fig:opt_pos_1}). \par
\begin{figure}
\begin{center}
\includegraphics[width=0.9\columnwidth]{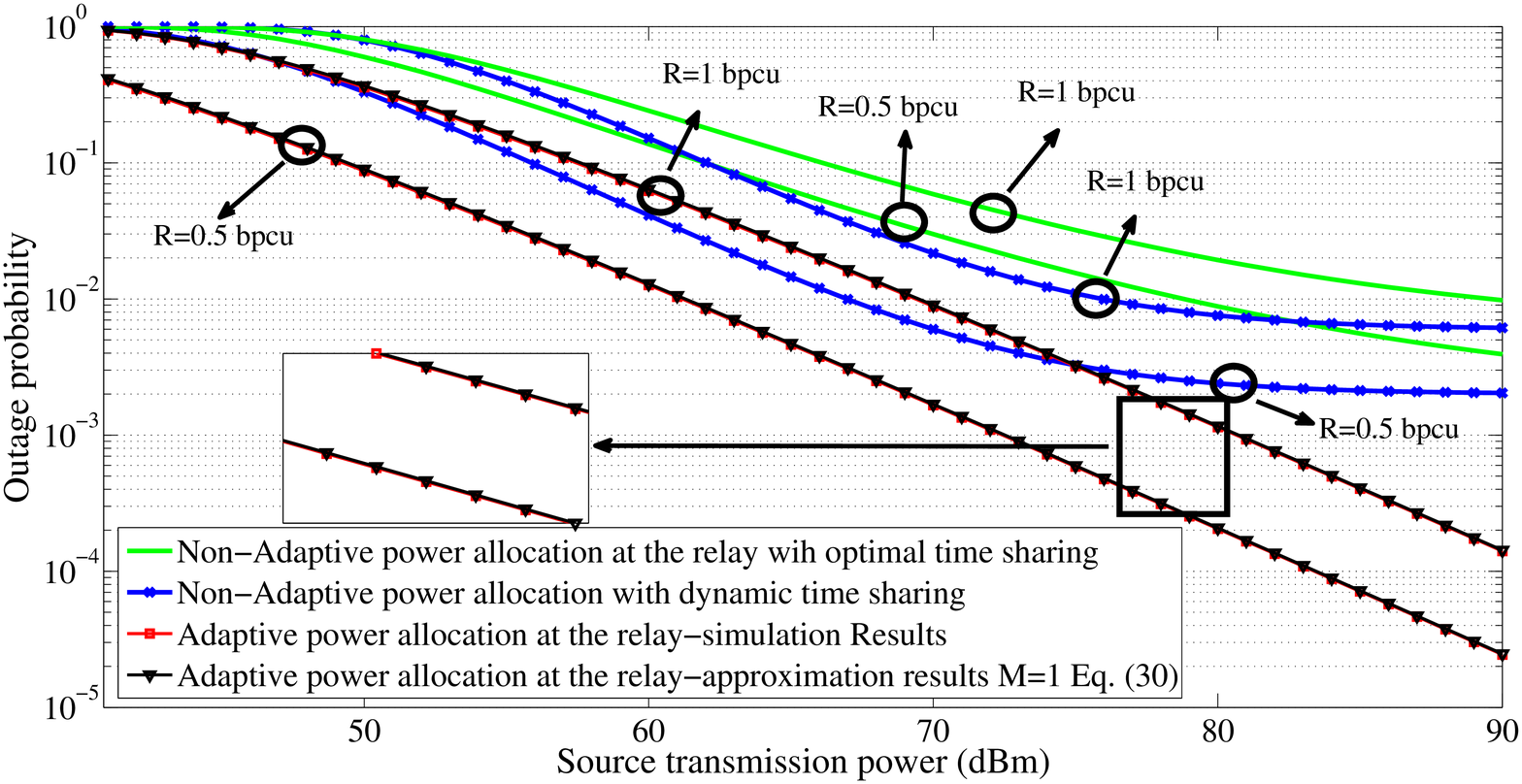}
\caption{Comparison of different protocols in term of outage probability, $\lambda_{\text{sr}} = \lambda_{\text{rd}}=1$, $E_{\text{proc}} = 23.01$ dBm,$P_{\text{active}} =
29.03 $ dBm,$ \nu = 2$, for the non-adaptive scheme $P_{\text{r}}^{\text{i}}=30$ dBm, for the adaptive scheme $ \theta= 0.4$.} \label{fig:out_dyn_fix}
\end{center}
\end{figure}

\begin{figure}
\begin{center}
\includegraphics[width=0.9\columnwidth]{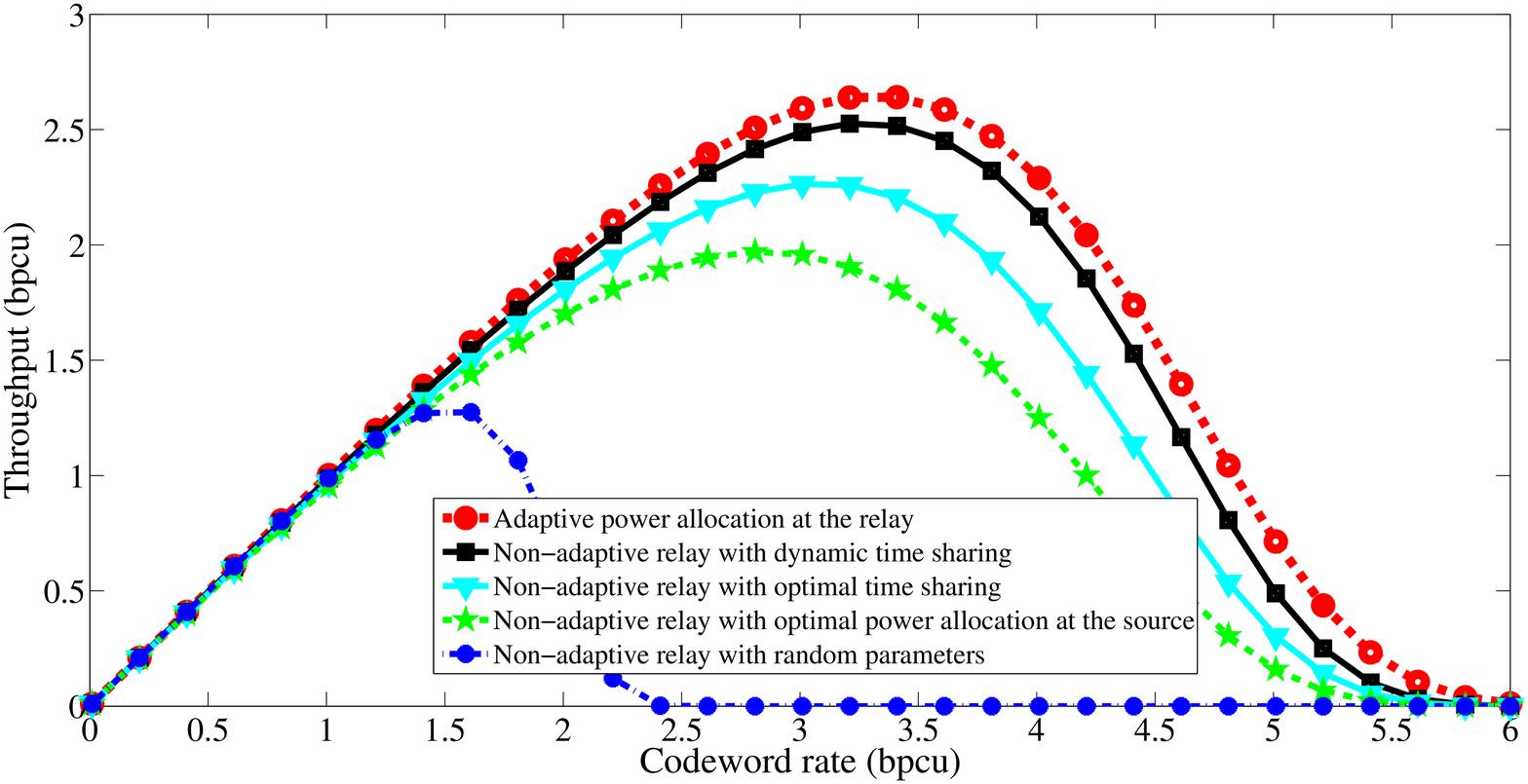}
\caption{Comparison of the different protocols in term of throughput,$\lambda_{\text{sr}} = \lambda_{\text{rd}}=1$, $E_{\text{proc}} = 23.01 $ dBm,$P_{\text{active}} =
29.03 $ dBm,$ \nu = 2$, for the non-adaptive scheme $P_{\text{r}}^{\text{i}}=30$ dBm, for the adaptive scheme $ \theta= 0.4$ and $P= 60$ dBm.} \label{fig:thr_dyn_fix}
\end{center}
\end{figure}

\textit{Outage Probability and Throughput of the Proposed Schemes for the Single-Relay Network:} Figures \ref{fig:out_dyn_fix} and \ref{fig:thr_dyn_fix} compare the performance of different schemes considered in this paper for the relay networks, in terms of outage probability and throughput, respectively. In Fig. \ref{fig:out_dyn_fix}, we have set the relay transmission power for non-adaptive schemes to $P_{\text{r}}^{\text{i}}=50$ dBm, the time sharing for the adaptive power allocation for the relay is set to $\theta=0.4$, and for the non-adaptive case with fixed time sharing is set to its optimal value given in Theorem \ref{thr:tet}. Moreover, Fig. \ref{fig:thr_dyn_fix} compares the throughput of different protocols. Here, the throughput of a given protocol with codeword rate $R$ and the outage probability of $\mathrm{Pr}\left(\text{outage}\right)$ is defined by
\begin{equation}
\mathcal{T}=R\left(1-\mathrm{Pr}\left(\text{outage}\right)\right).
\end{equation} From Fig. \ref{fig:out_dyn_fix}, at high-SNRs, the outage probability of the dynamic and fixed time sharing schemes are the same, since the outage probability of non-adaptive systems is mostly due to the rate outage in Phase 3, and it is independent of the source transmission power. However, at moderate SNRs, the dynamic time sharing outperforms the fixed time sharing, in terms of outage probability. Moreover, the relay with the adaptive power allocation outperforms the other scenarios, in terms of outage probability, and achieves the diversity order of one. Also, as seen in the figure, the approximation approach of (\ref{eq:ove_out_pro_adap})
is tight for a broad range of SNRs/parameter settings, such that two terms in (\ref{eq:ove_out_pro_adap}) provide accurate results.\par
Adaptive power allocation in the relay results in higher throughput compared to other protocols (Fig. 6). Also, setting time sharing or power allocation at the source to its optimal value significantly enhances the system throughput. Furthermore, with the parameter settings of the figure, the relay network with optimal time sharing, but uniform power allocation at the source, has higher throughput than the system with fixed time sharing, but optimal power allocation at the source. However, this is not a general conclusion and, depending on the parameter settings, different behaviors may be observed in the throughput.  \par
\begin{figure}
\begin{center}
\includegraphics[width=0.9\columnwidth]{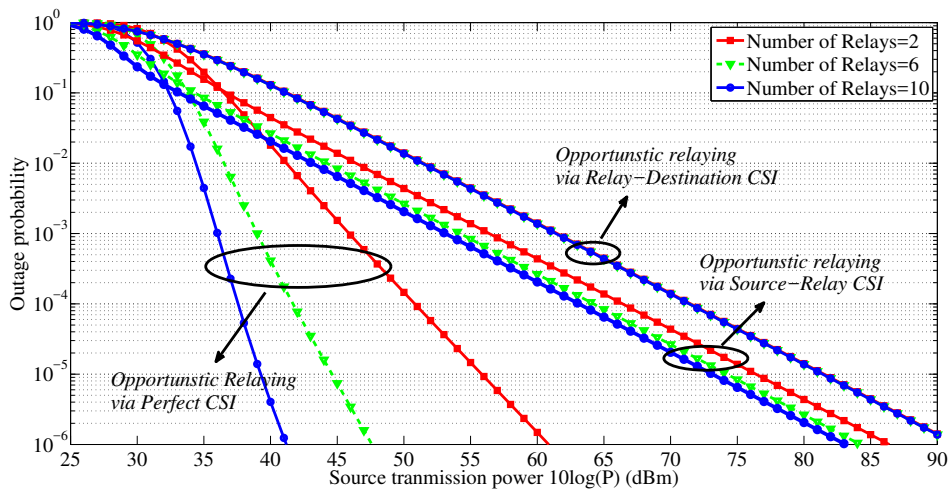}
\caption{Outage probability of multi-relay networks,  
$\lambda_{\text{sr}} = \lambda_{\text{rd}}=1$, $P_{\text{proc}} = 23.01 $ dBm , $P_{\text{active}} =
29.03 $ dBm , $ \nu = 2$
$R=0.5$  bpcu  and $\theta=0.4$.} \label{fig:mult_rel_1}
\end{center}
\end{figure}
\begin{figure}
\centering
\includegraphics[width=0.9\columnwidth]{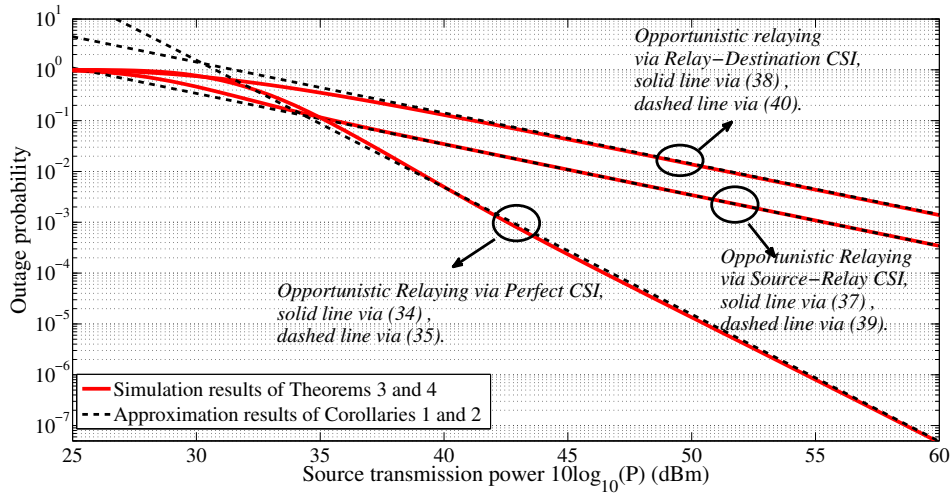}
\caption{Outage probability of multi-relay networks, 
$\lambda_{\text{sr}} = \lambda_{\text{rd}}=1$, $P_{\text{proc}} = 23.01 $ dBm , $P_{\text{active}} =29.03 $ dBm , $ \nu = 2$
$R=0.25$  bpcu , $\theta=0.4$ and $N=4$. Solid lines are the simulation results and dashed lines are the approximation results.} \label{fig:mult_rel_2}
\end{figure}
\begin{figure}
\centering
\includegraphics[width=0.9\columnwidth]{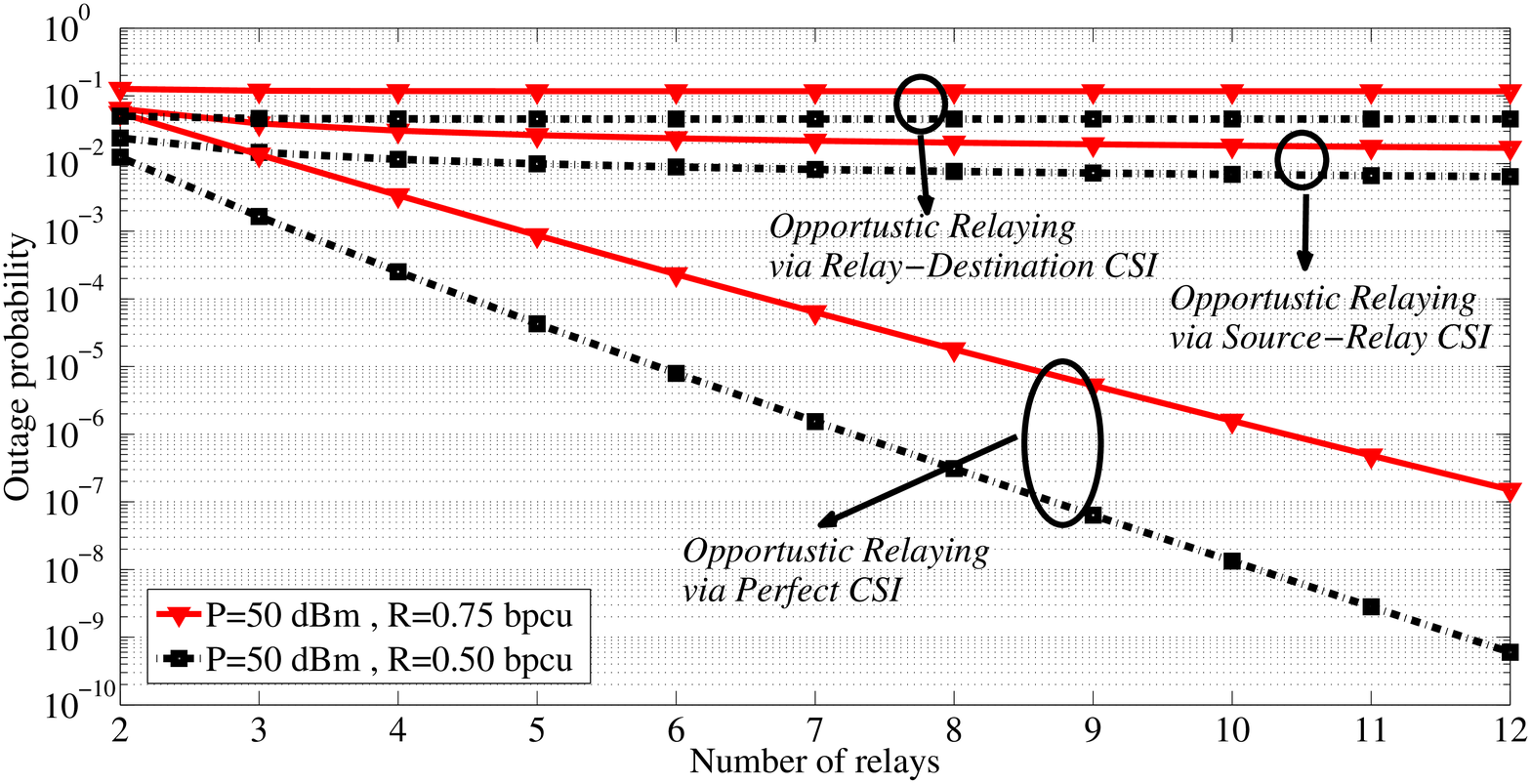}
\caption{Outage probability of multi-relay network vs Number of Relays, 
$\lambda_{\text{sr}} = \lambda_{\text{rd}}=1$, $P_{\text{proc}} = 23.01 $ dBm , $P_{\text{active}} =29.03 $ dBm , $ \nu = 2$, $\theta=0.4$.} \label{fig:mult_rel_3}
\end{figure}

\textit{On the Performance and Tightness of Approximations for the Multi-Relay Network:} Finally, Figs. \ref{fig:mult_rel_1}, \ref{fig:mult_rel_2}, and \ref{fig:mult_rel_3}  demonstrate the performance of the multi-relay scenario based on the selection criteria in (\ref{eq:opor_dual_hop}), (\ref{eq:opor_hop_1}) and (\ref{eq:opor_hop_2}).
Figure \ref{fig:mult_rel_1} shows the outage probability given in (\ref{eq:kokoli}), (\ref{eq:hop1_out}) and (\ref{eq:hop2_out}). Moreover, Fig. \ref{fig:mult_rel_2} demonstrates the accuracy of the high-SNR approximation of the outage probability given in Corollaries  \ref{corr:opor_full} and \ref{corr:opor_partial}. Then, Fig. \ref{fig:mult_rel_3} investigates the impact of increasing the number of relays on the outage probability.  From Fig. \ref{fig:mult_rel_1}, the opportunistic relay selection scheme based on the source-destination CSI, outperforms the relay selection based on the relay-destination CSI, as also stated in Corollary \ref{corr:opor_partial}. This is intuitively because the source-relay CSI not only affect the information transmission but also has impact on the energy transfer, whereas the relay-destination CSI has impact only on the information transmission in Phase 3. Also, both opportunistic relaying schemes with criteria (\ref{eq:opor_hop_1}) and (\ref{eq:opor_hop_2}) achieve the diversity order of one, independently of the number of relays. Moreover, from Fig. \ref{fig:mult_rel_2}, the high-SNR approximations of the outage probabilities presented in Corollaries  \ref{corr:opor_full} and \ref{corr:opor_partial} are tight for a broad range of SNRs. Finally, as seen in Fig. \ref{fig:mult_rel_3}, increasing the number of relays has a marginal influence on the performance of the relay selection with criteria (\ref{eq:opor_hop_1}) and (\ref{eq:opor_hop_2}), in comparison with (\ref{eq:opor_dual_hop}). 


\vspace{-1mm}

\vspace{-5mm}
\section{Conclusion} \label{sec:conc}
Considering imperfect power consumption models for the relay, we studied the outage probability and throughput of the relay networks with wireless energy and information transfer. We analyzed the system performance for two power allocation schemes of the relay, namely, non-adaptive and adaptive transmission powers. For the non-adaptive scheme, we derived the optimal time sharing, the optimal power allocation at the source, and the optimal relay position with an expected total energy consumption constraint such that the outage probability is minimized. Then, we extended our analysis to the multi-relay networks which incorporate opportunistic relaying. As demonstrated both analytically and numerically, the optimal power for the energy transfer signal increases linearly with the total energy of the source and decreases exponentially with the codeword rate. Then, in the multi-relay networks with $N$ relays, the outage probability of the opportunistic relaying increases with the inefficiency of the PA in power
of $\frac{N+1}{2}$ and $1$ for the cases with perfect-CSI and partial CSI, respectively. Finally, for multi-relay networks with partial CSI increasing the number of relays has marginal impact on the outage performance.
\appendix
\section{Proofs}
\subsection{Proof of Theorem \ref{thr:tet}}
Let $f_1(\theta)=\exp\left(-\frac{\gamma}{\lambda_{\text{rd}} P_{\text{r}}^{\text{i}}}\right)\exp\left(-\frac{(P_{\text{cons}}+P_{\text{proc}})\frac{\theta}{2}}{\lambda_{\text{sr}} P_{0}(1-\theta)}\right)$ and $f_2(\theta)=\exp\left(-\frac{\gamma}{\lambda_{\text{sr}} P_0}\right)\exp\left(-\frac{\gamma}{\lambda_{\text{rd}} P_{\text{r}}^{\text{i}}}\right)$. Also, we define $g_1(\theta)=\log(f_1(\theta))$ and $g_2(\theta)=\log(f_2(\theta))$. Taking the second derivative of $g_1(\theta)$, $g_2(\theta)$ functions, it is straightforward to show that they are concave functions in $\theta$. Therefore, the optimal time scheduling is found by setting the derivative of the objective functions equal to zero. Particularly, using $f_1(\theta)$, we have  
\begin{equation}\label{eq:opt_find_tet_1}
\begin{aligned}
\theta^*_1&=\underset{0 \leq \theta \leq 1}{\arg\max} \Bigg\{ g_1(\theta)=-\frac{\gamma}{\lambda_{\text{rd}} P_{\text{r}}^{\text{i}}} - \frac{(P_{\text{cons}}+P_{\text{proc}})\frac{\theta}{2} }{\lambda_{\text{sr}} P_0 (1-\theta)}\Bigg\}\\
&\stackrel{\text{(a)}}{=} \underset{0 \leq \theta \leq 1}{\arg} \Bigg\{\frac{P_{\text{cons}}+P_{\text{proc}}}{2\lambda_{\text{sr}}P_0(1-\theta)^2}=\frac{2R}{\lambda_{\text{rd}}P_{\text{r}}^{\text{i}}}\frac{\exp(\frac{2R}{\theta})}{\theta^2}\Bigg\}
\Rightarrow \theta^{*}_1=\frac{R}{R + \mathcal{W} \left(\frac{\sqrt{\lambda_{\text{rd}}P_{\text{r}}^{\text{i}}(P_{\text{cons}}+P_{\text{proc}})R}}{2\sqrt{\lambda_{\text{sr}}P_0}\exp(R)}\right)}.
\end{aligned}
\end{equation}
where $(\text{a})$ is obtained by setting $\frac{\text{d}g_1}{\text{d}\theta}=0$ and the last equality is obtained by some manipulations and the definition of Lambert $\mathcal{W}$ function \cite{onlambert}. For $f_2(\theta)$, we have
\begin{equation}\label{opt_p:ttet2}
\theta^*_2=\underset{0 \leq \theta \leq 1}{\arg\max}\Big\{g_2(\theta)=-\frac{\gamma}{\lambda_{\text{sr}} P_0} - \frac{\gamma}{\lambda_{\text{rd}} P_{\text{r}}^{\text{i}}}\Big\}.
\end{equation}
Since $g_2(\theta)$ is an increasing function of $\theta$, its minimum value is given by the boundary of the branches in (\ref{eq:ov_out_f}), i.e., 
\begin{equation} \label{eq:fixed_pp}
\begin{aligned}
\theta_2^*&=\underset{0 \leq \theta \leq 1}{\arg}\bigg\{\frac{\gamma}{\frac{\theta}{2}}= \frac{P_{\text{cons}}+P_{\text{proc}}}{(1-\theta)}\bigg\}
=\underset{0 \leq \theta \leq 1}{\arg}\left\lbrace \exp(\frac{2R}{\theta})-1 - \frac{P_{\text{cons}}+P_{\text{proc}}}{2}\frac{\theta}{1-\theta}=0\right\rbrace.
\end{aligned}
\end{equation}
Defining $y(\theta)=\exp(\frac{2R}{\theta})-1 - \frac{P_{\text{cons}}+P_{\text{proc}}}{2}\frac{\theta}{1-\theta}$, it is straightforward to show that $\lim_{\theta\to 0} y(\theta)=+\infty, \lim_{\theta \to 1} y(\theta)=-\infty$ and $y \left(\theta\right)$ is a  decreasing function of $\theta\in [0,1]$. Therefore, the solution of $y(\theta)=0$ will be unique for any values of $P_{\text{cons}}, P_{\text{proc}}, R$. Unfortunately, to the best of authors knowledge, there is no closed-form solution for $y(\theta)=0.$ Thus, considering the moderate/large values of $R$, $\theta^{*}_2$ is approximately given by
\begin{equation} \label{eq:approx_version_sol}
\begin{aligned}
&\theta_2^*\simeq\arg_\theta\bigg\{\exp\left(\frac{2R}{\theta}\right) - \frac{P_{\text{cons}}+P_{\text{proc}}}{2}\frac{\theta}{1-\theta}=0\bigg\} \\
&\Rightarrow \theta_2^* \simeq \frac{2R}{2R+ \mathcal{W} \left( R\left(P_{\text{cons}}+P_{\text{proc}}\right)\exp(-2R)\right)}.
\end{aligned}
\end{equation}
Note that branching condition in (\ref{eq:ov_out_f}) can be rephrased in term of $\theta$ as $\theta \in [\theta^{\star}_2,1]$ and $\theta \in [0,\theta^{\star}_2]$ for upper and lower branch, respectively. Let the $\frac{(1-\theta^{*}_1)\left(\exp\left(\frac{2R}{\theta^{*}_1}\right)-1\right)}{\frac{\theta^{*}_1}{2}} \geq P_{\text{cons}}+P_{\text{proc}}$. By calculating the first derivative $f_1(\theta)$ and $f_2(\theta)$, it can be proved that $f_1(\theta)$ is strictly decreasing function in $[\theta^{\star}_2,1]$ and $f_2(\theta)$ is strictly increasing in $[0,\theta^{\star}_2]$. Thus, the optimal $\theta$ for this case is $\theta^{\star}_2$ (lower branch of (\ref{eq:opt_thettttt})). In the other case where $\frac{(1-\theta^{*}_1)\left(\exp\left(\frac{2R}{\theta^{*}_1}\right)-1\right)}{\frac{\theta^{*}_1}{2}} \leq P_{\text{cons}}+P_{\text{proc}}$, we can use (\ref{eq:opt_find_tet_1}) to show that $\theta_1^*\in[\theta_2^*,1]$. In this case, $f_1(\theta)$ is increasing in $[\theta^{\star}_2,\theta^{\star}_1]$ and decreasing in $[\theta^{\star}_1,1]$. Also, $f_2(\theta)$ is strictly increasing in $[0,\theta^{\star}_2]$. Thus, the optimal value for $\theta$ is $\theta^{\star}_1$ (the upper branch of (\ref{eq:opt_thettttt})).
\vspace{-5mm}
\subsection{Proof of Theorem \ref{thr:pow}}
It is straightforward  to prove that the energy constraint of (\ref{opt_p:tet}) should hold with equality in the optimal case. Considering the equality, we can write
\begin{equation}\label{eq:rel_pse_psi}
P_{\text{s}}^{\text{i}}=\frac{P_{\text{max}}-P_{\text{s}}^{\text{e}}(1-\theta)}{\frac{\theta}{2} \exp\left(-\frac{1}{\lambda_{\text{sr}}}\frac{P_{\text{proc}}+P_{\text{cons}}}{P_{\text{s}}^{\text{e}}}\frac{\frac{\theta}{2}}{1-\theta}\right)}.
\end{equation}
Thus, the branching condition of (\ref{eq:ov_out_f}) is rephrased in term of $P_{\text{s}}^{\text{e}}$ as
\begin{equation}
\begin{aligned}
&\frac{\gamma}{\frac{P_{\text{max}}-P_{\text{s}}^{\text{e}}(1-\theta)}{\frac{\theta}{2} \exp\left(-\frac{1}{\lambda_{\text{sr}}}\frac{P_{\text{proc}}+P_{\text{cons}}}{P_{\text{s}}^{\text{e}}}\frac{\frac{\theta}{2}}{1-\theta}\right)} \frac{\theta}{2}} \leq \frac{P_{\text{cons}}+P_{\text{proc}}}{P_{\text{s}}^{\text{e}}(1-\theta)}\\
&\Rightarrow P_{\text{s}}^{\text{e}} \leq  
\frac{P_{\text{max}}\frac{\theta}{2\left(1-\theta\right)}\left(P_{\text{proc}}+P_{\text{cons}}\right)}{\lambda_{\text{sr}}P_{\text{max}}\mathcal{W}\left(\frac{\gamma\theta}{2\lambda_{\text{sr}}P_{\text{max}}}\exp\left(-\frac{\left(P_{\text{proc}}+P_{\text{cons}}\right)\frac{\theta}{2}}{\lambda_{\text{sr}}P_{\text{max}}}\right)\right)+\left(P_{\text{proc}}+P_{\text{cons}}\right)\frac{\theta}{2}}.
\end{aligned}
\end{equation}
Also, according to (\ref{eq:rel_pse_psi}), we have $P_{\text{s}}^{\text{e}} \leq \frac{P_{\text{max}}}{1-\theta}$, because $\exp\left(-\frac{\gamma}{\lambda_{\text{rd}} P_{\text{r}}^{\text{i}}}\right)\exp\left(-\frac{(P_{\text{cons}}+P_{\text{proc}}) \frac{\theta}{2}}{\lambda_{\text{sr}} P_{\text{s}}^{\text{e}}(1-\theta)}\right)$ is an increasing function of $P_s^e$, and the optimal value of  $P_s^e$ in the first branch of (\ref{eq:ov_out_f}) is given by
\begin{equation} \label{eq:alaaaki}
P_{\text{s}}^{\text{e}}=\min\left\lbrace \frac{P_{\text{max}}}{1-\theta},\frac{P_{\text{max}}\frac{\theta}{2\left(1-\theta\right)}\left(P_{\text{proc}}+P_{\text{cons}}\right)}{\lambda_{\text{sr}}P_{\text{max}}\mathcal{W}\left(\frac{\gamma\theta}{2\lambda_{\text{sr}}P_{\text{max}}}\exp\left(-\frac{\left(P_{\text{proc}}+P_{\text{cons}}\right)\frac{\theta}{2}}{\lambda_{\text{sr}}P_{\text{max}}}\right)\right)+\left(P_{\text{proc}}+P_{\text{cons}}\right)\frac{\theta}{2}}\right\rbrace.
\end{equation} 
Then, as $\mathcal{W}(x)>0$ for $x>0$, we have 
\begin{equation}\label{eq:alaaaki__}
\frac{P_{\text{max}}\frac{\theta}{2\left(1-\theta\right)}\left(P_{\text{proc}}+P_{\text{cons}}\right)}{\lambda_{\text{sr}}P_{\text{max}}\mathcal{W}\left(\frac{\gamma\theta}{2\lambda_{\text{sr}}P_{\text{max}}}\exp\left(-\frac{\left(P_{\text{proc}}+P_{\text{cons}}\right)\frac{\theta}{2}}{\lambda_{\text{sr}}P_{\text{max}}}\right)\right)+\left(P_{\text{proc}}+P_{\text{cons}}\right)\frac{\theta}{2}} \leq \frac{P_{\text{max}}}{1-\theta},
\end{equation}
and, from (\ref{eq:rel_pse_psi}), (\ref{eq:alaaaki}) and (\ref{eq:alaaaki__}), the optimal values of $P_{\text{s}}^{\text{e}}$ and $P_{\text{s}}^{\text{i}}$, in terms of (\ref{eq:ov_out_f}), are given by
\begin{equation}
\begin{aligned}\label{opt_point_1}
&(P_{\text{s}}^{\text{e}})^{\star}=\frac{P_{\text{max}}\frac{\theta}{2\left(1-\theta\right)}\left(P_{\text{proc}}+P_{\text{cons}}\right)}{\lambda_{\text{sr}}P_{\text{max}}\mathcal{W}\left(\frac{\gamma\theta}{2\lambda_{\text{sr}}P_{\text{max}}}\exp\left(-\frac{\left(P_{\text{proc}}+P_{\text{cons}}\right)\frac{\theta}{2}}{\lambda_{\text{sr}}P_{\text{max}}}\right)\right)+\left(P_{\text{proc}}+P_{\text{cons}}\right)\frac{\theta}{2}}\\
&(P_{\text{s}}^{\text{i}})^{\star} = \frac{P_{\text{max}}-(P_{\text{s}}^{\text{e}})^{\star}(1-\theta)}{\frac{\theta}{2} \exp\left(-\frac{1}{\lambda_{\text{sr}}}\frac{P_{\text{proc}}+P_{\text{cons}}}{(P_{\text{s}}^{\text{e}})^{\star}}\frac{\frac{\theta}{2}}{1-\theta}\right)},
\end{aligned}
\end{equation}
for the first branch of (\ref{eq:ov_out_f}). For the second branch of (\ref{eq:ov_out_f}), on the other hand, since the objective function $\exp\left(-\frac{\gamma}{\lambda_{\text{sr}} P_{\text{s}}^{\text{i}}}\right)\exp\left(-\frac{\gamma}{\lambda_{\text{rd}} P_{\text{r}}^{\text{i}}}\right)$ is a decreasing function of $P_\text{s}^\text{e}$, the optimal value of $P_\text{s}^\text{e}$ is given by the boundary of the feasible set which leads to the same value as (\ref{opt_point_1}). That is, the optimal point for $[(P^{\text{e}}_{\text{s}})^{\star}; (P^{\text{i}}_{\text{s}})^{\star}]$ is the same for both branches of the objective function. Thus, the outage-optimized power allocation rule is given by (\ref{opt_point_22}) as stated in the theorem.
\subsection{Proof of Theorem \ref{thr:opor_full}}
The selected relay's index, based on the criteria in (\ref{eq:opor_dual_hop}), is denoted by $i^*$.
Also, for the convenience, we define $\lambda_{1}=\frac{1}{\lambda_\text{sr}}$ and $\lambda_{2}=\frac{1}{\lambda_\text{rd}}$. Using Bayes' rule, we have  
\begin{equation} \label{eq:out_opor_asli}
\begin{aligned}  
\mathrm{Pr}\left(\text{outage}\right)
&=1-\underbrace{\mathrm{Pr}\left(g_{\text{sr}_{i^*}}\geq\max\{\frac{\alpha_1}{P_{\text{s}}^{\text{e}}},\frac{\alpha_2}{P_{\text{s}}^{\text{i}}}\},g_{\text{r}_{i^*\text{d}}}\left(g_{\text{sr}_{i^*}}-\frac{\alpha_1}{P_{\text{s}}^{\text{e}}}\right)\geq \frac{\alpha_3}{P_{\text{s}}^{\text{e}}}\big| g_{\text{sr}_{i^*}}\leq g_{\text{r}_{i^*\text{d}}} \right)}_{T_1}\mathrm{Pr}\left(g_{\text{sr}_{i^*}}\leq g_{\text{r}_{i^*\text{d}}}\right)\\& \ -\underbrace{\mathrm{Pr}\left(g_{\text{sr}_{i^*}}\geq\max\{\frac{\alpha_1}{P_{\text{s}}^{\text{e}}},\frac{\alpha_2}{P_{\text{s}}^{\text{i}}}\},g_{\text{r}_{i^*\text{d}}}\left(g_{\text{sr}_{i^*}}-\frac{\alpha_1}{P_{\text{s}}^{\text{e}}}\right)\geq \frac{\alpha_3}{P_{\text{s}}^{\text{e}}}\big| g_{\text{sr}_{i^*}}\geq g_{\text{r}_{i^*\text{d}}} \right)}_{T_2}\mathrm{Pr}\left(g_{\text{sr}_{i^*}}\geq g_{\text{r}_{i^*\text{d}}}\right).
\end{aligned}
\end{equation}
Let us define $v_i$ as a random variable which is equal to the channel gain at the relay $i$ which has smaller gain, i.e., $v_i=\min\{g_{\text{sr}_{i}},g_{\text{r}_{i\text{d}}}\}$, and, considering Rayleigh fading model, it follows the exponential distribution with rate $\lambda_1+\lambda_2$. Based on the selection criteria in (\ref{eq:opor_dual_hop}),  $v_{i^*}$'s PDF is
\begin{equation} \label{eq:dist_best}
f_{v_{i^*}}(z)=N\left(\lambda_1+\lambda_2\right)\exp\left(-\left(\lambda_1+\lambda_2\right)z\right)\left(1-\exp\left(-\left(\lambda_1+\lambda_2\right)z\right)\right)^{N-1},
\end{equation}
since, $v_{i^*}=\max\{v_{1},\hdots,v_{N}\}$\cite{pdf_ane}. In this way, $T_1$ in (\ref{eq:out_opor_asli}) is rephrased as 
\begin{equation}\label{eq:A1}
\begin{aligned}
T_1
&=\int\limits_{\max\{\frac{\alpha_1}{P_{\text{s}}^{\text{e}}},\frac{\alpha_2}{P_{\text{s}}^{\text{i}}}\}}^{\infty} \mathrm{Pr}\left(g_{\text{r}_{i^*\text{d}}} \geq \frac{\alpha_3}{\left(P_{\text{s}}^{\text{e}} x-\alpha_1\right)}\big| x\leq g_{\text{r}_{i^*\text{d}}}  \ , \ g_{\text{sr}_{i^*}}=x \right)f\left(x \bigg| x\leq g_{\text{r}_{i^*\text{d}}}\right)\text{d}x.
\end{aligned}
\end{equation}
In order to facilitate evaluation the integral in (\ref{eq:A1}), we define $\widetilde{\alpha}$ as the positive root of the  following equation
\begin{equation} \label{eq:dom}
\begin{aligned}
&x = \frac{\alpha_3}{P_{\text{s}}^{\text{e}}x-\alpha_1} \Rightarrow P_{\text{s}}^{\text{e}}x^2-\alpha_1x-\alpha_3=0\\
 &\Rightarrow \widetilde{\alpha}\triangleq\frac{\alpha_1+\sqrt{\alpha_1^2+4P_{\text{s}}^{\text{e}}\alpha_3}}{2P_{\text{s}}^{\text{e}}}.
\end{aligned}
\end{equation} 
Also, note that $\widetilde{\alpha} \geq \frac{\alpha_1}{P_{\text{s}}^{\text{e}}}$.
Here, considering (\ref{eq:A1}), (\ref{eq:dom}) and  the case $\frac{\alpha_1}{P_{\text{s}}^{\text{e}}}\geq \frac{\alpha_2}{P_{\text{s}}^{\text{i}}}$, $T_1$ is obtained as
\begin{equation}\label{eq:A1_}
\begin{aligned}
T_1&=\int\limits_{\frac{\alpha_1}{P_{\text{s}}^{\text{e}}}}^{\widetilde{\alpha}}\frac{\exp\left(-\frac{\lambda_2 \alpha_3}{P_{\text{s}}^{\text{e}}x-\alpha_1}\right)}{\exp\left(-\lambda_2 x\right)}f\left(x \big| x\leq g_{\text{r}_{i^*\text{d}}}\right)\text{d}x + \int\limits_{\widetilde{\alpha}}^{\infty}f\left(x \big| x\leq g_{\text{r}_{i^*\text{d}}}\right)\text{d}x\\
&\stackrel{\text{(b)}}{=}\int\limits_{\frac{\alpha_1}{P_{\text{s}}^{\text{e}}}}^{\widetilde{\alpha}}\frac{\exp\left(-\frac{\lambda_2 \alpha_3}{P_{\text{s}}^{\text{e}}x-\alpha_1}\right)}{\exp\left(-\lambda_2 x\right)}f_{v_{i^*}}\left(x\right)\text{d}x + \int\limits_{\widetilde{\alpha}}^{\infty}f_{v_{i^*}}\left(x\right)\text{d}x\\
&=N\left(\lambda_1+\lambda_2\right)\sum\limits_{k=0}^{N-1} {N-1 \choose k}(-1)^k\int\limits_{\frac{\alpha_1}{P_{\text{s}}^{\text{e}}}}^{\widetilde{\alpha}}J_{k}\left(x\right)\text{d}x+1-\left(1-\exp\left(-\left(\lambda_1+\lambda_2\right)\widetilde{\alpha}\right)\right)^N,
\end{aligned}
\end{equation}
where $J_{k}\left(x\right)\triangleq\exp\left(-\frac{\lambda_2 \alpha_3}{P_{\text{s}}^{\text{e}}x-\alpha_1}-(k+1)\lambda_1x -k\lambda_2x\right)$, and (b) is based on the fact that $f\left(x \big| x\leq g_{\text{r}_{i^*\text{d}}}\right)=f_{v_{i^*}}(x)$ where $f_{v_{i^*}}(x)$ is given in (\ref{eq:dist_best}). 
Considering $\frac{\alpha_1}{P_{\text{s}}^{\text{e}}}\geq \frac{\alpha_2}{P_{\text{s}}^{\text{i}}}$, $T_2$ in (\ref{eq:out_opor_asli}) is given by 
\begin{equation}\label{eq:A2_}
\begin{aligned}
T_2&= \int\limits_{0}^{\infty} \mathrm{Pr}\left(g_{\text{sr}_{i^*}}\geq \frac{\alpha_1}{P_{\text{s}}^{\text{e}}},g_{\text{sr}_{i^*}}\geq \frac{\alpha_3}{P_{\text{s}}^{\text{e}}y}+\frac{\alpha_1}{P_{\text{s}}^{\text{e}}}\big| g_{\text{sr}_{i^*}}\geq y,g_{\text{r}_{i^*\text{d}}}=y\right)f\left(y\big|g_{\text{sr}_{i^*}}\geq y\right)\text{d}y\\
&\stackrel{\text{(c)}}{=}\int \limits_{0}^{\widetilde{\alpha}} \mathrm{Pr}\left(g_{\text{sr}_{i^*}}\geq \frac{\alpha_3}{P_{\text{s}}^{\text{e}}y}+\frac{\alpha_1}{P_{\text{s}}^{\text{e}}}\big| g_{\text{sr}_{i^*}}\geq y,g_{\text{r}_{i^*\text{d}}}=y\right)f_{v_{i^*}}\left(y\right)\text{d}y+\int \limits_{\widetilde{\alpha}}^{\infty}f_{v_{i^*}}\left(y\right)\text{d}y \\
&=N\left(\lambda_1+\lambda_2\right)\sum\limits_{k=0}^{N-1} {N-1 \choose k}(-1)^k\int\limits_{0}^{\widetilde{\alpha}}U_{k}\left(y\right)\text{d}y+1-\left(1-\exp\left(-\left(\lambda_1+\lambda_2\right)\widetilde{\alpha}\right)\right)^N,
\end{aligned}
\end{equation}
where $U_{k}\left(y\right) \triangleq \exp\left(-\lambda_1 \left(\frac{\alpha_3}{P_{\text{s}}^{\text{e}}y}+\frac{\alpha_1}{P_{\text{s}}^{\text{e}}}\right)-k\lambda_1y-(k+1)\lambda_2y\right)$ and (c) is found by the fact that $f\left(x \big| x\leq g_{\text{r}_{i^*\text{d}}}\right)=f_{v_{i^*}}(x)$ with $f_{v_{i^*}}(x)$ given in (\ref{eq:dist_best}).\\
On the other hand, if $\frac{\alpha_1}{P_{\text{s}}^{\text{e}}}\leq \frac{\alpha_2}{P_{\text{s}}^{\text{i}}}$, following the same procedure as (\ref{eq:A1_}) and (\ref{eq:A2__}), $T_1$ and $T_2$ are found as 
\begin{equation} \label{eq:A1__} 
T_1=\begin{cases}
1-\left(1-\exp\left(-\left(\lambda_1+\lambda_2\right)\frac{\alpha_2}{P_{\text{s}}^{\text{i}}}\right)\right)^N  &\text{ $\frac{\alpha_2}{P_{\text{s}}^{\text{i}}}\geq \widetilde{\alpha} \  \& \ \frac{\alpha_2}{P_{\text{s}}^{\text{i}}} \geq \frac{\alpha_1}{P_{\text{s}}^{\text{e}}}  $}\\ 
N\left(\lambda_1+\lambda_2\right)\sum\limits_{k=0}^{N-1} {N-1 \choose k}(-1)^k\int\limits_{\frac{\alpha_2}{P_{\text{s}}^{\text{i}}}}^{\widetilde{\alpha}}J_{k}\left(x\right)\text{d}x+1-\left(1-\exp\left(-\left(\lambda_1+\lambda_2\right)\widetilde{\alpha}\right)\right)^N &\text{ $\frac{\alpha_2}{P_{\text{s}}^{\text{i}}}\leq \widetilde{\alpha} \  \& \ \frac{\alpha_2}{P_{\text{s}}^{\text{i}}} \geq \frac{\alpha_1}{P_{\text{s}}^{\text{e}}}$}
\end{cases}.
\end{equation}
\begin{equation} \label{eq:A2__}
T_2=\begin{cases}
1-\left(1-\exp\left(-\left(\lambda_1+\lambda_2\right)\frac{\alpha_2}{P_{\text{s}}^{\text{i}}}\right)\right)^N  &\text{ $\frac{\alpha_2}{P_{\text{s}}^{\text{i}}}\geq \widetilde{\alpha} \  \& \ \frac{\alpha_2}{P_{\text{s}}^{\text{i}}} \geq \frac{\alpha_1}{P_{\text{s}}^{\text{e}}} $}\\ 
N\left(\lambda_1+\lambda_2\right)\sum\limits_{k=0}^{N-1} {N-1 \choose k}(-1)^k\int\limits_{\frac{\alpha_2}{P_{\text{s}}^{\text{i}}}}^{\widetilde{\alpha}}U_{k}\left(y\right)\text{d}y+1-\left(1-\exp\left(-\left(\lambda_1+\lambda_2\right)\widetilde{\alpha}\right)\right)^N &\text{ $\frac{\alpha_2}{P_{\text{s}}^{\text{i}}}\leq \widetilde{\alpha} \  \& \ \frac{\alpha_2}{P_{\text{s}}^{\text{i}}} \geq \frac{\alpha_1}{P_{\text{s}}^{\text{e}}}$}
\end{cases}.
\end{equation}
Moreover, it is straightforward to prove that $\mathrm{Pr}\left(g_{\text{sr}_{i^*}}\geq g_{\text{r}_{i^*\text{d}}}\right)=1-\mathrm{Pr}\left(g_{\text{sr}_{i^*}}\leq g_{\text{r}_{i^*\text{d}}}\right)=\frac{\lambda_2}{\lambda_1+\lambda_2}$. Finally, substituting (\ref{eq:A1_}), (\ref{eq:A2_}),  (\ref{eq:A1__}), and (\ref{eq:A2__}) into (\ref{eq:out_opor_asli}) the outage probability expression in (\ref{eq:kokoli}) yields.
\subsection{Proof of Proposition \ref{corr:opor_full}}
For the notational convenience, we define $\lambda_1=\frac{1}{\lambda_{\text{sr}}}$, $\lambda_2=\frac{1}{\lambda_{\text{rd}}}$ and $\omega=\sqrt{\frac{1}{P}}$. We analyze the outage probability of the first branch in (\ref{eq:kokoli}), then the analysis is readily extended to other branches in (\ref{eq:kokoli}). The first term of the first branch in (\ref{eq:kokoli}) is rephrased as
\begin{equation}\label{eq:mohem2}
\begin{aligned}
\left(1-e^{-\left(\lambda_1+\lambda_2\right)\widetilde{\alpha}}\right)^N&=\sum\limits_{k=0}^{N}{N \choose k} (-1)^k e^{-k\left(\lambda_1+\lambda_2\right)\widetilde{\alpha}}\\
&\stackrel{\text{(d)}}{=}\sum\limits_{i=0}^{\infty}\underbrace{\left[\sum\limits_{k=0}^{N}{N \choose k}(-1)^k k^i\right]}_{S_i}\frac{\left[-\left(\lambda_1+\lambda_2\right)\widetilde{\alpha}\right]^i}{i!}\\
&\stackrel{\text{(e)}}{=}\sum\limits_{i=N}^{\infty}\left[\sum\limits_{k=0}^{N}{N \choose k}(-1)^k k^i\right]\frac{\left(-\left(\lambda_1+\lambda_2\right)\widetilde{\alpha}\right)^i}{i!}\\
&=\left(\left(\lambda_1+\lambda_2\right)\sqrt{\alpha_3}\right)^N \omega^N+\frac{S_{N+1}}{(N+1)!}\left(-\left(\lambda_1+\lambda_2\right)\sqrt{\alpha_3}\right)^{N+1}\omega^{N+1}+o\left(\omega^{N+1}\right).
\end{aligned}
\end{equation}
Here, (d) is obtained by the Taylor series expansion of $e^{-k\left(\lambda_1+\lambda_2\right)}$ and (e) is found by using the  properties of sum of binomial coefficients \cite[Eq. 0.154]{tabebesel} which 
\begin{equation}\label{prop:bio}
\begin{aligned}
&\sum\limits_{k=0}^{N}{N \choose k}(-1)^k k^i=0 \quad \text{for} \quad  0\leq i<N,
&\sum\limits_{k=0}^{N}{N \choose k}(-1)^k k^N=(-1)^NN!.
\end{aligned}
\end{equation}
Also, $f\left(x\right)=o\left(g\left(x\right)\right)$ is defined as $\lim_{x \to 0} \frac{f\left(x\right)}{g\left(x\right)}=0$. Furthermore, the second term in the first branch of (\ref{eq:kokoli}) is rephrased as  
\begin{equation}\label{eq:div_opor1}
\begin{aligned}
&N\lambda_1\sum\limits_{k=0}^{N-1} {N-1 \choose k}(-1)^k \int\limits_{\alpha_1 \omega^2}^{\widetilde{\alpha}}J_k\left(x\right)\text{d}x =N\lambda_1\sum\limits_{k=0}^{N-1} {N-1 \choose k}(-1)^k \int\limits_{\alpha_1 \omega_2}^{\widetilde{\alpha}}e^{-\left(\lambda_1+k\left(\lambda_1+\lambda_2\right)\right) x}q(x)\text{d}x\\
&\stackrel{\text{(f)}}{=}N\lambda_1\sum\limits_{k=0}^{N-1} {N-1 \choose k}(-1)^k\int\limits_{\alpha_1 \omega^2}^{\widetilde{\alpha}}\sum\limits_{i=0}^{\infty}\frac{(-1)^i\left(\lambda_1+k\left(\lambda_1+\lambda_2\right)\right)^ix^i}{i!}q(x)\text{d}x\\
&=N\lambda_1\int\limits_{\alpha_1 \omega^2}^{\widetilde{\alpha}}\sum\limits_{i=0}^{\infty}\frac{(-1)^ix^i\lambda_1^i}{i!}\sum\limits_{l=0}^{i}{i \choose l}\underbrace{\left(\frac{\lambda_1+\lambda_2}{\lambda_1}\right)^l\left[\sum\limits_{k=0}^{N-1}{N-1 \choose k}(-1)^kk^l\right]}_{\Phi_l}q(x)\text{d}x\\
&=N\lambda_1\int\limits_{\alpha_1 \omega^2}^{\widetilde{\alpha}}\sum\limits_{i=N-1}^{\infty}\frac{(-1)^ix^i\lambda_1^i}{i!}\sum\limits_{l=0}^{i}{i \choose l}\Phi_lq(x)\text{d}x \triangleq \mathcal{B}_1,
\end{aligned}
\end{equation}
where $q(x)\triangleq e^{-\frac{\lambda_2\alpha_3}{P\left(x-\frac{\alpha_1}{P}\right)}}$ and (f) is obtained by the Taylor series expansion. Then, the last step follows from $\Phi_l=0$ for $0 \leq l < N-1$ (\ref{prop:bio}). On the other hand, 
\begin{equation}\label{eq:order_}
\begin{aligned}
| \mathcal{B}_1 |&\stackrel{\text{(g)}}{\leq}\bigg| N\lambda_1\int\limits_{\alpha_1 \omega^2}^{\widetilde{\alpha}}\sum\limits_{i=N-1}^{\infty}\frac{(-1)^ix^i\lambda_1^i}{i!}\sum\limits_{l=0}^{i}{i \choose l}\Phi_l\text{d}x\bigg|\\&=I_{N}\omega^{N}+I_{N+1}\omega^{N+1}+o(\omega^{N+1}),
\end{aligned}
\end{equation}
where, (g) comes from $|q(x)|<1$ and the triangle inequality.
Thus, $I_{N}$ and $I_{N+1}$ characterize the asymptotic behavior of $\mathcal{B}_1$.
For the sake of determining $I_{N}$ and $I_{N+1}$, we use the power series expansion of $q(x)$ which is given by
\begin{equation}\label{eq:taghrib_q}
q\left(x\right)=e^{-\frac{\lambda_2\alpha_3}{P\left(x-\frac{\alpha_1}{P}\right)}}=\sum\limits_{j=0}^{\infty} \left(-\frac{\lambda_2\alpha_3}{\alpha_1}\sum\limits_{n=1}^{\infty}\left(\frac{Px}{\alpha_1}\right)^{-n}\right)^j=\sum\limits_{m=0}^{\infty}\Gamma_m  \left(\frac{x}{\omega^{2}}\right)^{-m},
\end{equation}
where $\Gamma_m$ denotes the coefficient of $x^{-m}$. Substituting (\ref{eq:taghrib_q}) into (\ref{eq:div_opor1}), $\mathcal{B}_1$ is found as
\begin{equation} \label{eq:opor_cor1}
\begin{aligned}
\mathcal{B}_1&=N\lambda_1\int\limits_{\alpha_1 \omega^2}^{\widetilde{\alpha}}\sum\limits_{m=0}^{\infty}\sum\limits_{i=N-1}^{\infty}\Gamma_m \frac{(-1)^ix^{i-m}\lambda_1^i}{i! \omega^{-2m} }\sum\limits_{l=0}^{i}{i \choose l}\Phi_l  \text{d}x
 = I_{N}\omega^N+I_{N+1}\omega^{N+1}+o\left(\omega^{N+1}\right).
\end{aligned}
\end{equation}
Thus, $I_{N}$ and $I_{N+1}$ are found by considering $(i=N-1, m=0)$ and $(i=N,m=0)\&(i=N-1,m=1)$, respectively. After some algebraic manipulations, we get 
\begin{equation}\label{eq:opor_cor1_}
\begin{aligned}
I_{N}&=\lambda_1\left(\lambda_1+\lambda_2\right)^{N-1}\alpha_3^{\frac{N}{2}},\\
I_{N+1}&=\left[\frac{N\left(-1\right)^N \left(\lambda_1\right)^{N+1}}{\left(N+1\right)!}\left( N\Phi_{N-1}+\Phi_{N}\right)-\frac{N\lambda_1\lambda_2 \left(\lambda_1+\lambda_2\right)^{N-1}}{N-1}\right]\left(\sqrt{\alpha_3}\right)^{N+1}.
\end{aligned}
\end{equation}
  Following the same steps as in (\ref{eq:div_opor1}), (\ref{eq:order_}), (\ref{eq:taghrib_q}), (\ref{eq:opor_cor1}), and (\ref{eq:opor_cor1_}) yields
\begin{equation}\label{eq:opor_cor2}
\begin{aligned}
 \mathcal{B}_2&\triangleq N\lambda_2\sum\limits_{k=0}^{N-1} {N-1 \choose k}(-1)^k \int\limits_{0}^{\widetilde{\alpha}}U_k\left(x\right)\text{d}x
 = V_N\omega^N+V_{N+1}\omega^{N+1}+o\left(\omega^{N+1}\right),
\end{aligned}
\end{equation}
where 
\begin{equation}\label{eq:opor_cor2_}
\begin{aligned}
V_{N}&=\lambda_2\left(\lambda_1+\lambda_2\right)^{N-1}\alpha_3^{\frac{N}{2}},\\
V_{N+1}&=\left[\frac{N\left(-1\right)^N \left(\lambda_2\right)^{N+1}}{\left(N+1\right)!}\left( N\Phi_{N-1}+\Phi_{N}\right)-\frac{N\lambda_1\lambda_2\left(\lambda_1+\lambda_2\right)^{N-1}}{N-1}\right]\left(\sqrt{\alpha_3}\right)^{N+1}.
\end{aligned}
\end{equation}
Finally, substituting (\ref{eq:mohem2}), (\ref{eq:opor_cor1}), (\ref{eq:opor_cor1_}),  (\ref{eq:opor_cor2}) and (\ref{eq:opor_cor2_}) into the first branch in (\ref{eq:kokoli}) results in
\begin{equation}\label{eq:wer}
\begin{aligned}
&\left(1-e^{-\left(\lambda_1+\lambda_2\right)\widetilde{\alpha}}\right)^N -\left(N\sum\limits_{k=0}^{N-1} {N-1 \choose k}(-1)^k\left(\lambda_1\int\limits_{\alpha_1\omega^2}^{\widetilde{\alpha}}J_k\left(x\right)\text{d}x+\lambda_2\int\limits_{0}^{\widetilde{\alpha}}U_k\left(x\right)\text{d}x\right)\right)\\
&\simeq\Omega_N \left(\frac{\gamma\nu \theta }{2(1-\theta)}\right)^{\frac{N+1}{2}} \omega^{N+1}=\Omega_N \left(\frac{\gamma\nu \theta }{2(1-\theta)}\right)^{\frac{N+1}{2}} (\frac{1}{P})^{\frac{N+1}{2}}
\end{aligned}
\end{equation}
where $\Omega_N$ is defined in (\ref{eq:muuu}). Also, the above analysis can be applied readily to the last branch of (\ref{eq:kokoli}) and the same results as (\ref{eq:wer}) yields. Also, note that at high SNRs, $\alpha_2 \omega^2 \leq \widetilde{\alpha}$. Hence, either first or third branching condition holds in the outage probability in (\ref{eq:kokoli}).
\subsection{Proof of Proposition \ref{corr:opor_partial} }
For the case that the relay is selected based on the first hop, the outage is given by (\ref{eq:hop1_out}). At high SNRs, the lower limit of the integral is approximately equal to zero. Thus, we have
\begin{equation}\label{eq:hop1_hsnr}
\begin{aligned}
\mathrm{Pr}\left(\text{outage}\right)^{\text{SR-CSI}}&\simeq 1-\sum\limits_{i=0}^{N-1}\frac{N}{\lambda_{\text{sr}}}(-1)^i {N-1 \choose i}e^{-\frac{(i+1)\alpha_1 }{\lambda_{\text{sr}}P}}  \int\limits_{0}^{\infty} e^{-\frac{ \alpha_3 }{\lambda_{\text{rd}}Px}}e^{-\frac{(i+1)x}{\lambda_{\text{sr}}}}\text{d}x \\
&\stackrel{\text{(h)}}{=}1-N\sum\limits_{i=0}^{N-1} {N-1 \choose i} 
\frac{(-1)^i}{i+1} e^{-\frac{(i+1)\alpha_1 }{\lambda_{\text{sr}}P}} \sqrt{ \frac{4\alpha_3 (i+1)}{\lambda_{\text{sr}} \lambda_{\text{rd}}P}}\text{K}_1\left(\sqrt{ \frac{4\alpha_3 (i+1)}{\lambda_{\text{sr}} \lambda_{\text{rd}}P}}\right)\\
&\stackrel{\text{(i)}}{=}1-N\sum\limits_{i=0}^{N-1} {N-1 \choose i} 
\frac{(-1)^i}{i+1}e^{-\frac{(i+1)\alpha_1 }{\lambda_{\text{sr}}P}} \left(1+\frac{\alpha_3 (i+1)}{\lambda_{\text{sr}} \lambda_{\text{rd}}P}\log\left(\frac{\alpha_3 (i+1)}{\lambda_{\text{sr}} \lambda_{\text{rd}}P}\right)\right)\\
&\stackrel{\text{(j)}}{=}1-N\sum\limits_{i=0}^{N-1} {N-1 \choose i} 
\frac{(-1)^i}{i+1} \bigg(1+\frac{\alpha_3 (i+1)}{\lambda_{\text{sr}} \lambda_{\text{rd}}P}\log\left(\frac{\alpha_3 (i+1)}{\lambda_{\text{sr}} \lambda_{\text{rd}}}\right)\\
&\hspace{5.45cm}-\frac{\alpha_3 (i+1)}{\lambda_{\text{sr}} \lambda_{\text{rd}}P}\log(P) -\frac{(i+1)\alpha_1 }{\lambda_{\text{sr}}P}+o(\frac{1}{P})\bigg)\\
&\simeq\frac{N\alpha_3}{\lambda_{\text{sr}}\lambda_{\text{rd}}} \sum\limits_{i=0}^{N-1} {N-1 \choose i} (-1)^{i+1} \log(i+1) \frac{1}{P}.
\end{aligned}
\end{equation}
Here, (h) is obtained by the definition of the modified Bessel function. For the equation in (i), we use the tight of approximation of $t\text{K}_1(t)\simeq 1+\frac{t^2}{2}\log\left(\frac{t}{2}\right)$ for $t\to 0$\cite[Eq. 24]{poor}, (j) is found by considering the first order Taylor series of $e^{-(i+1)\lambda_1\alpha_1 t}$, and finally,in the last step, we use the properties of the sum of binomial coefficient  
\begin{align}
&\sum_{i=0}^{N-1}\limits {N-1 \choose i} (-1)^i \frac{1}{i+1} =\frac{1}{N},\label{eq:bio1}\\
&\sum_{i=0}^{N-1} \limits {N-1 \choose i} (-1)^i =0.\label{eq:bio2}
\end{align}
When the relay selection is performed based on the relay-destination channel, the outage is given by (\ref{eq:hop2_out}). Considering $\alpha_1>\alpha_2$, we have
\begin{equation}\label{eq:opor_2__}
\begin{aligned}
\mathrm{Pr}\left(\text{outage}\right)^{\text{RD-CSI}}&= 1+  \exp\left(\frac{-\alpha_1}{\lambda_{\text{sr}}P}\right) \sum\limits_{i=1}^{N} {N \choose i} (-1)^i  \sqrt{\frac{4 i \alpha_3  }{\lambda_{\text{sr}}\lambda_{\text{rd}}P}}\text{K}_1\left(\sqrt{\frac{4 i \alpha_3  }{\lambda_{\text{sr}}\lambda_{\text{rd}}}P}\right)\\
&\stackrel{\text{(k)}}{\simeq}1+\left(1-\frac{\alpha_1}{\lambda_{\text{sr}}P}\right)\sum\limits_{i=1}^{N}{N \choose i}(-1)^i \left(1+\frac{i \alpha_3  }{\lambda_{\text{sr}}\lambda_{\text{rd}}P} \log\left(\frac{i \alpha_3  }{\lambda_{\text{sr}}\lambda_{\text{rd}}P} \right)\right)+o(\frac{1}{P})\\
&\stackrel{\text{(l)}}{=}\frac{\alpha_1}{\lambda_{\text{sr}}P}+\frac{\alpha_3}{\lambda_{\text{sr}}\lambda_{\text{rd}}P}\sum\limits_{i=1}^{N}{N \choose i}(-1)^ii\log\left(i\right)\\
&=\frac{\alpha_1}{\lambda_{\text{sr}}P}+\frac{N\alpha_3}{\lambda_{\text{sr}}\lambda_{\text{rd}}P}\sum\limits_{i=0}^{N-1}{N-1 \choose i}(-1)^{i+1}\log\left(i+1\right),
\end{aligned}
\end{equation}
where, (k) is obtained by following the similar procedure in (\ref{eq:hop1_hsnr}) and for (l), we use (\ref{prop:bio}), (\ref{eq:bio2}). Then, the last equation follows from the property of binomial coefficient $i{N \choose i}=N{N-1 \choose i-1}$ which results the upper branch of (\ref{eq:hop2_high_snr}). \\ 
For the case $\alpha_2 \geq \alpha_1$, the outage at high SNRs is approximately given by
\begin{equation}
\mathrm{Pr}\left(\text{outage}\right)^{\text{RD-CSI}}\simeq 1+\sum\limits_{i=1}^{N}(-1)^i {N \choose i}\frac{1}{\lambda_{\text{sr}}} e^{-\frac{\alpha_2}{\lambda_{\text{sr}}P}}\int\limits_{0}^{\infty}e^{-\frac{i \alpha_3}{\lambda_{\text{rd}}Px}}e^{-\frac{x}{\lambda_{\text{sr}}}}\text{d}x,
\end{equation}
and the high SNRs approximation can be obtained by following the same line of arguments as in (\ref{eq:opor_2__}) that results the lower branch of (\ref{eq:hop2_high_snr}).
\small{
\bibliographystyle{IEEEtran} 
\bibliography{ref_new}}
\end{document}